\documentclass[10pt,pre,aps,twocolumn,showpacs,superscriptaddress,floatfix]{revtex4-1}

\usepackage{graphicx}
\usepackage{amsmath}
\usepackage{amssymb}
\usepackage{wasysym}
\usepackage{subfigure}
\usepackage{color}
\usepackage{amsmath} 
\begin{document}

\title{Bridging lattice-scale physics and continuum field theory \\ with quantum Monte Carlo simulations} 

\author{Ribhu K. Kaul}
\affiliation{Department of Physics and Astronomy, University of Kentucky, Lexington, Kentucky 40506, USA}

\author{Roger G. Melko}
\affiliation{Department of Physics and Astronomy, University of Waterloo, Ontario, N2L 3G1, Canada, and \\
Perimeter Institute for Theoretical Physics, Waterloo, Ontario N2L 2Y5, Canada}

\author{Anders W. Sandvik}
\affiliation{Department of Physics, Boston University, 590 Commonwealth Avenue, Boston, Massachusetts 02215, USA}

\begin{abstract}
We discuss designer Hamiltonians---lattice models tailored to be free from sign problems (``de-signed'') when simulated with quantum 
Monte Carlo methods but which still host complex  many-body states and quantum phase transitions of interest in condensed matter 
physics. We focus on quantum spin systems in which competing interactions lead to non-magnetic ground states. These states and the 
associated quantum phase transitions can be studied in great detail, enabling direct access to universal properties and connections 
with low-energy effective quantum field theories. As specific examples, we discuss the transition from a N\'eel antiferromagnet to 
either a uniform quantum paramagnet or a spontaneously symmetry-broken valence-bond solid in SU($2$) and SU($N$) invariant spin models. 
We also discuss anisotropic (XXZ) systems harboring topological Z$_2$ spin liquids and the XY$^*$ transition. We briefly review recent 
progress on quantum Monte Carlo algorithms, including ground state projection in the valence-bond basis and direct computation 
of the Renyi variants of the entanglement entropy.
\end{abstract}

\maketitle

\tableofcontents

\date{\today}

\section{Introduction}
\label{sec:intro}

Understanding ground states of interacting quantum systems is one of the central defining challenges of condensed matter physics. On the one
hand, quantum field theory is a powerful and general framework for studies of low-energy properties of complex, strongly correlated and entangled 
quantum matter \cite{Sachdev11}. On the other hand, this continuum approach has potential pitfalls due to its generality, and limitations due to its 
insensitivity to lattice scale physics, making it important to also carry out direct studies of microscopic lattice Hamiltonians. A fruitful direction 
is to combine the advantages of both methods by analyzing data from unbiased numerical investigations of lattice Hamiltonians with quantum field 
theoretic predictions of emergent low energy properties. Such connections between the two approaches can have significant synergistic effects in advancing 
our understanding of quantum many-body phenomena---as they have had and continue to have in classical statistical mechanics, especially in studies 
of thermal phase transitions \cite{cardy1988:fss,Chaikin00}. 

Computational many-body research faces important challenges since the quantum models of interest are often beyond the reach 
of existing algorithms. Two-dimensional (2D) and 3D frustrated quantum spin systems and fermion systems are the main groups 
of difficult systems, since they suffer from the infamous ``sign problem'' in quantum Monte Carlo (QMC) simulations
\cite{Loh90,Henelius00,Nyfeler08}. The sign problem arises when the weight function in the configuration space constructed using the Euclidean path 
integral or other mappings to an effective statistical-mechanics problem is not positive definite, hence invalidating the interpretation of the weights as a probability distribution for importance sampling. The approach we advocate and review here is to construct particular 
``designer Hamiltonians'', which host interesting ground states and quantum phase transitions but still are amenable to 
large-scale QMC studies without sign problems (i.e., they are designed and ``de-signed'' to be practically useful). 

A common misconception is that interesting designer Hamiltonians should not exist---if a model is sign-problem 
free it must be trivial or uninteresting. In this article, we will review some recent counter-examples to this pessimistic view, 
from the area of quantum magnetism. 
Moreover, advances in QMC 
methods \cite{Sandvik91,Evertz93,Beard96,WormA,Sandvik99,Sandvik10a} have made it possible to truly approach the 
low-energy limit of many important models without any approximations (beyond controllable statistical errors), and to compute 
quantities of current interest in condensed matter physics and quantum information theory, e.g., the fidelity 
\cite{Schwandt09}, the geometric tensor \cite{Degrandi11}, and the entanglement entropy \cite{Hastings10, Melko10}. We will 
briefly address some of these technical developments as well. We first make some further remarks on field theory and the philosophy 
of the synergistic use of designer Hamiltonians, and then outline the topics covered in this Review.

\subsection{Field theory and numerical studies of Hamiltonians}

The natural starting point for theoretical models of condensed matter systems are microscopic Hamiltonians.
An effective field theoretic description  of a given Hamiltonian 
is constructed based on a combination of insights and assumptions on the properties of the Hamiltonian that control the long distance physics according to the renormalization group (RG) approach. These properties most famously include symmetry and dimensionality, but more subtle effects like the nature of the topological defects also play an important role.  The fixed points of the field theory control the long distance physics.
The RG approach may result in flows to a previously known fixed point, or to some fixed point
whose properties are not known. The RG itself often relies on approximations that are not easy
to justify, and, if the flow is to some unknown state, it is challenging to characterize it. 
Often to control the RG flows, analytic expansions around some special simplifying limits are considered. Well known examples are the $\epsilon$ expansion
around the upper critical dimension and the $1/N$ expansion in the number of components $N$ (spin components, flavors of fermions, etc.)
\cite{herbut2007:book,barber1972:lgNeps}. The
extrapolation to the $\epsilon$ or $N$ of interest is often problematic, however, because only low-order
expansions are feasible in practice---or worse still, the expansion may not converge at the value of $\epsilon$ or $N$ of primary interest.

One example of the above approach of deriving effective field theories from microscopic Hamiltonians by a limiting expansion is the $(d+1)$-dimensional
nonlinear $\sigma$-model description of interacting quantum spins in $d$ dimensions, which was derived by Haldane in 
the semi-classical limit of large spin $S$ \cite{Haldane83,Chakravarty89,Auerbach94}.  Another example---central to this review article---is the 
non-compact CP$^{N-1}$ description of the intriguing  N\'eel to valence-bond solid (VBS) ``deconfined'' quantum-critical point in 2D 
SU($N$) quantum magnets, which can be justified in the large-$N$ limit \cite{Senthil04a,Sachdev08}. Whether or not the theory remains correct down to 
small $N$, in particular $N=2$, remains a challenging question.

In view of the uncertainties with analytic methods highlighted above, it is important to test the predictions of field theories in some unbiased way,
starting from a microscopic description of the system or phenomenon of interest. In some cases there are exact solutions, 
e.g., the Bethe Ansatz solution of the $S=1/2$ Heisenberg chain \cite{Bethe31} has been crucial for testing 
the non-linear $\sigma$-model description of this class of critical spin chains. The exact AKLT state \cite{affleck88} for a special 
version of the $S=1$ chain was similarly important in confirming Haldane's conjecture of the qualitative differences between 
half-odd integer and integer spins \cite{Haldane83}. Exact solutions are rare, however, and mostly limited to 1D models.
 
Another, more general approach is to study Hamiltonians numerically. Exact diagonalization of the Hamiltonian is possible 
only for very small systems; currently up to $42$ $S=1/2$ spins \cite{Nakano11,Lauchli11}. White's density matrix 
renormalization (DMRG) scheme \cite{White92} and related approaches based on matrix-product states \cite{Schollwock05} 
have enabled more detailed studies of the low-energy physics of 1D systems, including also ``ladders'' of several coupled chains
\cite{Dagotto96}. QMC studies are also competitive in 1D in the absence of sign problems \cite{Sandvik04,Jeckelmann02,Mund09}.

While there has been some progress for 2D systems with DMRG \cite{Stoudenmire12} and tensor-product \cite{Murg09,Gu08,Xie09} states, which are
higher-dimensional generalizations of the  matrix-product states generated in DMRG \cite{Rommer97}, in general these methods have not yet reached the point 
where one  can obtain unbiased results for generic nontrivial models. There are encouraging developments, however, of calculations addressing challenging systems and 
approaching the level of accuracy where definite conclusions can be drawn \cite{Yan11,Bauer12}.

Presently, however, the only numerical approach with which one can routinely reach sufficiently large 2D and 3D lattices in a completely unbiased way 
is QMC simulations \cite{Evertz03,Sandvik10b}; but, as already noted, they are restricted to systems free from sign problems. The class of models 
for which the sign problem is either absent or evadable
still contains a vast range of Hamiltonians with non-trivial and interesting 
ground states and quantum phase transitions. To study low-energy emergent properties, the interactions do not necessarily have to correspond in detail to 
any particular real material---although some times that is also possible \cite{Zvyagin07}. The idea is to design a sign-problem free Hamiltonian in such a way that it contains 
a particular macroscopic (low-energy) phenomenon of interest---so that universal physics is captured. 

Bench-mark results obtained from such designer Hamiltonians, representing various physical phenomena in a prototypical manner, can be very useful experimentally. 
Here our main focus will be on the theory side. With unbiased large-scale QMC simulations, one can test field theories proposed to capture specific classes of 
quantum many-body states or quantum phase transitions. Moreover, explorations of designer Hamiltonians can also serve as ``experiments'' for discovering 
novel phenomena, and, thereby, stimulate further theoretical developments.

A designer Hamiltonian corresponds to an effective classical statistical-mechanics problem (with a real action), which arises out of the construction 
of the sampling space in the QMC scheme implemented. In some cases quantum phenomena in $d$ dimensions can be studied using Monte Carlo simulations with 
the simplest generalization to a classical model with the same global symmetries in $d+1$ dimensions \cite{Rieger94,Sorensen92,Nahum11} (e.g., the thermal 
phase transition of the 2D classical Ising model is in the same universality class as the quantum phase transition of the transverse-field Ising chain 
\cite{Suzuki76,Sachdev11}). Often, however, the effective classical model resulting from the construction of a QMC configuration space, in a complete treatment
of a given a quantum Hamiltonian, has unusual degrees of freedom and interactions, unanticipated in the study of classical statistical mechanics. The naive 
mapping with just $d \to d+1$ can then miss important and intriguing phenomena special to quantum mechanics \cite{Senthil04a,Fradkin04,Sachdev08}. Thus, in our 
view the approach of directly studying quantum mechanical designer Hamiltonians is an important direction in forming an unbiased picture of quantum 
criticality and emergent phenomena.

\subsection{Outline of the Review}

In this Review, we will focus on some recent examples where it has been possible to study quantum phase transitions in detail 
by QMC simulations and to make quantitative comparisons of extracted universal quantities with field theoretic results. We also discuss other fascinating quantum phenomena, e.g., emergent gauge fields and entanglement, where QMC studies are
playing a critical role in establishing benchmarks and gaining insights that are hard or impossible to extract from the field theories.

In Sec.~\ref{sec:methods} we discuss the sign problem and some of the QMC schemes used in large-scale studies of quantum spin systems, 
including treatments of SU($N$) symmetry and recent progress in computing the Renyi entanglement entropy. In Sec.~\ref{sec:su2models} 
we give examples of conventional and deconfined quantum-critical points in SU(2) invariant models, and in Sec.~\ref{sec:sunmodels} 
we show how generalization to SU($N$) symmetry can be used to directly connect to large-$N$ studies within field theory. In Sec.~\ref{sec:u1models} 
we discuss spin liquid states and associated quantum phase transitions in U($1$) symmetric, highly frustrated ``XXZ'' spin models (or, equivalently, 
interacting hard-core bosons). We conclude in Sec.~\ref{sec:discussion} with some further discussion and an outlook on future prospects and challenges.

\section{Modern QMC methods}
\label{sec:methods}

There is a wide range of QMC methods available for studies of different classes of lattice models \cite{Assaad07,Evertz03,Sandvik10b}. In this Review we 
focus on spin models, and therefore restrict the discussion to the types of methods most suitable for them. These methods are also often well suited for studies 
of bosons, where a substantial body of work has been carried out in the past several years, stimulated by ultracold atoms in optical 
lattices \cite{Kashurnikov02,Wessel04,Kato08,Pollet10}. For fermions, there are alternative approaches, based upon auxiliary fields decoupling the interactions and 
subsequent analytical tracing-out of the fermions \cite{Hirsch82,Assaad07}. These methods have a much less severe fermion sign problem (in some cases completely 
avoiding it, e.g., for particle-hole symmetric systems)---with the exception of 1D systems, where the methods discussed here also work well (since the signs 
from anticommutation normally do not appear in this geometry). We will only give a brief non-techical overview of QMC methods here, with references to 
more detailed accounts.

\subsection{The sign problem}
\label{ss:sign}

Solving many-body quantum mechanical problems on a computer is {\em generally} exponentially hard (in computing time versus system size) 
with known algorithms. The statistical mechanics of a $d$-dimensional quantum Hamiltonian can be rewritten in terms of some classical variables 
$C$ in $d+1$ dimensions, i.e., the partition function ($T>0$) or the normalization ($T=0$) can be cast into the form
\begin{equation}
\label{eq:wc}
Z=\sum_C W_C.
\end{equation}
Systems of interest can, for practical, known ways of transforming into this form, have both positive and negative weights $W_C$, 
invalidating the usual interpretation of $W_C/Z$ as a probability distribution in Monte Carlo simulations. Formally, this difficulty 
can be side-stepped by not including the sign in the probability, using $P_C \propto |W_C|$, and compensating for this by weighting 
observables with the sign;
\begin{eqnarray}
\langle O \rangle & = & \frac{\sum_C W_C O_C}{\sum_C W_C} = \frac{\sum_C O_C {\rm Sign}(W_C) |W_C|}{\sum_C {\rm Sign}(W_C)|W_C|} \nonumber\\
& = & \frac{\langle {\rm Sign} \cdot O\rangle_{|W|}}{\langle {\rm Sign}\rangle_{|W|}}.
\label{expvalue}
\end{eqnarray}
The problem here is that the numerator and denominator in the last expression both approach zero
exponentially in system size and inverse temperature, and, hence, resolving the ratio within 
statistical noise is exponentially hard. The advantage of importance sampling (time speed-up from exponential 
to polynomial scaling) is then lost. This predicament is the QMC sign problem \cite{Loh90,Henelius00,Nyfeler08}.

Here, we consider classes of quantum Hamiltonians for which it is possible to choose a basis in which all the $W_C\geq 0$. 
These sign-problem free models can generally be simulated in polynomial time by 
importance sampling, allowing access to large system sizes (with the exception of ``glassy'' systems and
other systems where even classical Monte Carlo simulations scale exponentially). In some cases, sign problems that
appear intractable at first-sight can be solved in more sophisticated ways, e.g., in the Meron algorithm \cite{Chandrasekharan99}. 
Here we implicitly consider cases where the ``de-signing'' is essentially automatic, being a direct consequence of the non-positivity of the  
off-diagonal matrix elements of the Hamiltonian.
 
\subsection{$T>0$ and $T=0$ QMC methods}
\label{ss:method}

In {\it finite-temperature} methods, the goal is to compute thermal averages
\begin{equation}
\langle A\rangle = \frac{1}{Z}{\bf Tr}\{ A{\rm e}^{-\beta H}\},~~~
Z = {\bf Tr}\{ {\rm e}^{-\beta H}\},
\label{za1}
\end{equation}
where $\beta=1/T$. As an alternative to studying the ground state in the limit of  $T\to 0$, in a {\it ground-state projector} method some 
operator $P(\beta)$  is applied to a ``trial state'' $|\Psi_0\rangle$, such that $|\Psi_\beta \rangle = P(\beta)|\Psi_0\rangle$ approaches the 
ground state when $\beta \to \infty$. An expectation value
\begin{equation}
\langle A\rangle = \frac{1}{Z}\langle \Psi_\beta|A|\Psi_\beta\rangle,~~~~ Z = \langle \Psi_\beta|\Psi_\beta\rangle,
\label{za2}
\end{equation}
tends to the true ground state expectation value, $\langle A\rangle$ $\to$ $\langle 0| A|0\rangle$. For the projector, one can use the imaginary-time
evolution operator $P(\beta)={\rm e}^{-\beta H}$ or a high power of the Hamiltonian; $P(m)=H^m$. Here $m \propto \beta N$ gives the same rate of 
convergence for the two choices of projectors for a given system size $N$. This follows from a Taylor expansion of the time evolution operator, which for large 
$\beta$ is dominated by powers of the order $n=\beta |E_0|$, where $E_0$ is the ground state energy.

\subsection{Path integrals and stochastic series expansions}

The task is now to rewrite $Z$ in Eqs.~(\ref{za1}) or (\ref{za2}) in the form (\ref{eq:wc}), and expectation values as Eq.~(\ref{expvalue}), with classical variables 
without diagonalizing $H$. In practice, $T=0$ and $T>0$ schemes for a given model are often very similar. In both cases, the exponential 
operator can be treated with path-integral methods---world lines in discrete \cite{Suzuki77,Hirsch82} or continuous \cite{Beard96,Prokofev96,Prokofev98} 
imaginary time---starting from a product of ``time slice'' evolution operators,
\begin{equation}
{\rm e}^{-\beta H}= \prod_{i=1}^M {\rm e}^{-\Delta H},~~~~ \Delta=\frac{\beta}{M}.
\label{slicing}
\end{equation}
Complete sets of states are then inserted between the exponentials (and the trace is further taken at $T>0$). For small $\Delta$, the
matrix elements of the exponentials can be evaluated approximately, giving the form (\ref{eq:wc}) with $W_C$ a product of $M$ matrix elements. 
In modern methods the limit $\Delta \to 0$ is taken at the algorithmic level \cite{Prokofev96,WormA,Beard96} and the Monte Carlo sampling is of 
paths in continuous imaginary time. Older approaches employed the Suzuki-Trotter decomposition \cite{Suzuki76} in Eq.~(\ref{slicing}), which 
typically leads to an error $\mathcal{O}(\Delta^2)$.

An alternative to Eq.~(\ref{slicing}) is to start from a series expansion \cite{Handscomb62,Sandvik91},
\begin{equation}
{\rm e}^{-\beta H}= \sum_{n=0}^\infty \frac{\beta^n}{n}(-H)^n,
\label{series}
\end{equation}
and insert complete sets of states between each instance of $H$. The power $n$ itself is importance-sampled along with the terms of $H$ 
({\it Stochastic Series Expansion}; SSE). This expansion is sharply peaked around $\langle n\rangle = -\beta E$, where $E$ is the total internal energy 
(under the assumption that there is no sign problem), which leads to a computational effort scaling as $\beta N$, which is the same as in the 
path-integral approaches. Again, $W_C$ in Eq.~(\ref{eq:wc}) is a product of matrix elements. 

The continuous-time path integral can also be seen as a variant of SSE, where the evolution operator is written in the interaction picture, 
with $H=H_0+V$ and expanding only in the perturbation $V$ to some conveniently chosen $H_0$ \cite{Prokofev96,Sandvik97a}. The full SSE, where $H_0=0$, $V=H$, 
is normally more efficient for spin systems, while the continuous-time variant should be better for certain boson systems \cite{Troyer03}.

Regardless of how the exponential operator is treated, the difference between Eqs.~(\ref{za1}) and (\ref{za2}) is essentially only in the boundary 
conditions in the imaginary time direction; periodic for $T>0$ (due to the trace being taken) and dictated by the nature of the trial state in $T=0$ 
projections. An equal superposition over all basis states corresponds to fully open boundaries, while other choices of the trial state leads to ``biased''
boundaries with some weighting of the bra and ket states (depending on the details of the trial state). The projector approach with $P(\beta)={\rm e^{-\beta H}}$
is often called the Path Integral Ground State method (PIGS) in the context of continuous-space systems \cite{Sarsa00,Vitali08}.

There is no sign problem in these approaches for spin systems with non-positive definite off-diagonal matrix elements (for a properly chosen trial state in the 
case of the $T=0$ approach). For bipartite antiferromagnetic interactions a sublattice rotation can accomplish this, or, equivalently, one can note that
for such models the signs are ``invisible'', because all non-vanishing terms of the path integral or series expansion have an even number of off-diagonal 
matrix elements. For further details of sign issues, see, e.g., Refs.~\cite{Henelius00,Evertz03,Nyfeler08}. Note that there are no restrictions on the diagonal 
interactions in the chosen basis, so that one can study, e.g., spin and boson systems in which the potential energy is highly frustrated (as we will discuss below 
in Sec.~\ref{sec:u1models}).

\subsection{Sampling with loop and cluster algorithms}

A breakthrough leading to today's efficient QMC algorithms was the realization by Evertz {\it et al.}~\cite{Evertz93} that the classical Swendsen-Wang (SW) cluster 
Monte Carlo method \cite{Swendsen88} could be generalized to loop algorithms (i.e., generating clusters in the form of loops) for vertex models, and that these models 
are very similar to the effective statistical mechanics problems arising in the transformations of quantum problems discussed above. While the loop algorithm has 
been adapted to many different models, both within the path-integral \cite{Kawashima94,Beard96,Harada02,Evertz03} and SSE \cite{Sandvik99} frameworks, they cannot 
be applied to all cases (in analogy with the SW method, which also has limited applicability). The loop concept has also been further generalized, 
however, to ``worms'' \cite{Prokofev96,Prokofev98, WormA} and ``directed loops'' \cite{Sandvik99,Syljuasen02}, which are essentially loops that can self-intersect 
during their construction and incorporate the detailed-balance principle in a more general manner. Recently such 
an algorithm was also formulated without detailed balance \cite{Suwa10}. In addition, for some models where loop algorithms and their generalizations are not 
applicable (e.g., transverse-field Ising models), other efficient generalizations of the SW cluster concept have been developed \cite{Rieger99,Sandvik03}.

\subsection{SU(2)-symmetric models}
\label{su2method}

\begin{figure}
\includegraphics[width=7.0cm, clip]{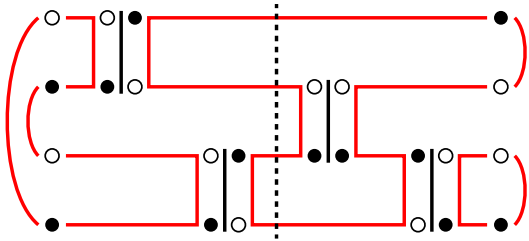}
\caption{A configuration in a valence-bond projector simulation of a 4-site Heisenberg chain; here with a small projection power, 
$m=2$ \cite{Sandvik10a}. The vertical bars indicate singlet-projectors operators, while the arcs on the left (bra) and right (ket) represent 
valence bond configurations of the sampled trial state. Open and closed circles indicate up and down spins, which are unchanged between 
operators and represented by horizontal lines there. These lines, along with the valence bonds, form loops that can be flipped (i.e., 
all spins traversed by the loop are flipped) without changing the configuration weight. A configuration in a $T>0$ simulation corresponds 
to imposing periodic boundary conditions in the ``time'' direction (reflecting the trace operation), instead of the independent boundaries 
terminated by valence bonds.}
\label{loops}
\end{figure}

A particularly convenient class of trial states in projector QMC studies of SU($2$) invariant interactions is the amplitude-product states in the overcomplete
valence-bond (singlet pair) basis \cite{Liang88}, which were recently generalized to include bond correlations \cite{Lin12}. These total-singlet states incorporate 
Marshall's sign rule in a convenient way and also automatically have the appropriate momentum for the ground state of a periodic system (thereby from the outset 
filtering out a significant fraction of the excited states). The Hamiltonian in this case can typically be expressed in terms of 
two-spin singlet-projector operators  (individual ones or products of two or more of them);
\begin{equation}
P_{ij}=\hbox{$\frac{1}{4}$}-{\bf S}_i\cdot {\bf S}_j = |s_{ij}\rangle\langle s_{ij}|,
\label{psinglet}
\end{equation}
where $s_{ij}$ denotes the singlet state of spins $i$ and $j$. A power of the Hamiltonian $H^m$, for fixed $m$ or in a series expansion, Eq.~(\ref{series}),
is expanded into all possible products of these projectors, which act on the ket trial state in (\ref{za2}). In early versions of this scheme, the resulting 
paths of valence bond states were sampled directly \cite{Liang90,Santoro99,Sandvik05}, while in more recent formulations the diagonal and off-diagonal parts of the 
projectors are also sampled, along with spin configurations compatible with the valence-bond configurations of the trial state (which are also sampled) 
\cite{Sandvik10c}. An example of a configuration in such a simulation is shown in Fig.~\ref{loops}. This configuration space is amenable to the very efficient 
loop updates discussed above \cite{Sandvik10a}. Apart from the boundary conditions in the ``propagation'' direction (and the associated sampling of the
valence bonds in the trial state), this kind of projector method is in practice implemented in a very similar way as the SSE method for SU($2$) 
models \cite{Sandvik10b}.

Evaluation of expectation values (``measurements'') in projector QMC simulations are carried out at the mid-point of the configuration (while in $T>0$
methods averages can be taken over the whole time range, due to the time-periodicity), as indicated 
in Fig.~\ref{loops}. In the valence-bond basis, when propagating the bra and ket states in the pure valence-bond basis (which corresponds to averaging 
over all compatible spin configurations), most quantities of interest can be expressed using the loops of the transition graph formed when superimposing 
the projected bra and ket bond configurations \cite{Liang88,Sutherland88}. Estimators for operators $A$ expressed in the space $\{\mathcal{L}\}$ of 
transition-graph loops are typically of the form
\begin{equation}
A(\mathcal{L}) = \frac{\langle V_l|A|V_r\rangle}{\langle V_l|V_r \rangle},
\label{aloop}
\end{equation}
where the overlap of the two valence-bond states is given by
\begin{equation}
\langle V_l|V_r\rangle = 2^{N_\circ - N/2},
\label{vboverlap}
\end{equation}
where $N_\circ$ is the number of loops and $N$ the system size. The numerator in Eq.~(\ref{aloop}) depends on the loop structure, 
e.g., for $A={\bf S}_i \cdot {\bf S}_j$, the estimator $A(\mathcal{L})$ is $0$ if sites $i$ and $j$ belong to different loops and $\pm 3/4$ for 
sites in the same loop (with $+$ and $-$ for the same and different sublattices, respectively). Higher-order correlation functions are discussed 
in Refs.~\cite{Beach06} and \cite{Tang11b}.

\subsection {Generalizations from SU(2) to SU(N)}
\label{ss:su2N}

The above methods for SU($2$) models can be extended to a certain class of spin models with an SU($N$) symmetry on bipartite lattices. First, note again 
that the SU(2) Heisenberg operator is simply a singlet projector, Eq.~(\ref{psinglet}). In order to generalize this operator to SU($N$), we choose models 
that have an SU($N$) spin with a fundamental representation on the A sub-lattice and an SU($N$) spin with a conjugate to the fundamental representation 
on the B sub-lattice~\cite{affleck1985:lgN,Read89}. Denoting the $N$ states on each site by $|\alpha\rangle$, where $1\leq \alpha \leq N$, 
and the generators of the fundamental representation by $T^a$ ($N\times N$ matrices), the SU($N$) singlet of a spin $i$ on the A sub-lattice and $j$ 
on the B sub-lattice is given by, 
\begin{equation}
|s^N_{ij}\rangle = \frac{1}{\sqrt{N}}\sum_\alpha |\alpha_i\alpha_j\rangle. 
\end{equation}
We note here that for $N=2$ and on bipartite lattices, this singlet is equivalent to the standard way of writing the singlet as 
$(|\uparrow \downarrow\rangle -|\downarrow \uparrow\rangle)/\sqrt{2}$, by making the transformation $|\uparrow \rangle\rightarrow |\downarrow\rangle, |\downarrow \rangle\rightarrow -|\uparrow\rangle$ on one of the sublattices. We can use  $|s^N_{ij}\rangle$ to rewrite the manifestly SU($N$) invariant version of Eq.~(\ref{psinglet}) as an explicitly sign-problem 
free projection operator,
\begin{equation}
\sum_a T^a_i \cdot
T^{*a}_j + \frac{1}{N^2} = |s^N_{ij}\rangle \langle s^N_{ij}|\equiv P_{ij},
\end{equation}
generalizing the Heisenberg interaction for two spins on {\em opposite} sublattices. 

Another interaction of interest is an SU($N$) invariant interaction between sites having the same representation; the permutation operator, 
which is defined in the following way by its action on a ket:
\begin{equation}
\frac{1}{N}\Pi_{ij}|\alpha_i\beta_j\rangle \equiv |\beta_i\alpha_j\rangle. 
\label{pija}
\end{equation}
We can now relate this operator to the generators $T^a$:
\begin{equation}
\sum_a T^a_i \cdot
T^{a}_j +\frac{1}{N^2} =\Pi_{ij}.
\label{pijb}
\end{equation}
This is an SU($N$) invariant generalization of the SU($2$) Heisenberg interaction for two spins on the {\em same} sublattice.

Hamiltonians based the SU($N$) operators $P_{ij}$ and $\Pi_{ij}$ are sign-problem free if they come with negative signs, in which case
they generalize the SU($2$) antiferromagnetic and ferromagnetic Heisenberg exchange, respectively. In terms of the definitions of $P_{ij}$ and 
$\Pi_{ij}$, the algorithms presented in Sec.~\ref{ss:method} can be simply generalized from SU(2) to SU($N$) by extending the number of ``colors'' 
(of the spins and the loops) from 2 to $N$~\cite{harada2003:sun,beach2009:sun,kaul2011:j1j2}. As a consequence of this, estimators expressed in terms
of transition-graph loops are also modified, e.g., in the state overlap Eq.~(\ref{vboverlap}) $2$ is replaced by $N$. One can even generalize
simulations in the pure valence-bond basis to non-integer $N$ \cite{beach2009:sun}.

\subsection{Renyi entropies via the replica trick} 
\label{ss:renyi}

In addition to conventional physical observables evaluated according to Eqs.~(\ref{za1}) and (\ref{za2}), e.g., various correlation functions, 
QMC methods have recently been developed that are capable of measuring the degree of {\it entanglement} in a quantum system \cite{EntangleMeasure}. Several 
quantities related to entanglement have been explored in the QMC context for their abilities to identify and characterize quantum phases and phase 
transitions \cite{Tommaso1,Tommaso2,fluc1,fluc2,fid1,fid2}.  In this Review, we concentrate of measures of the entanglement entropy, specifically 
the {\it Renyi} entropies \cite{renyi},
\begin{equation}
S_{\alpha} = \frac{1}{1-\alpha} \ln \Big[ {\rm Tr}\big( \rho_A^{\alpha} \big) \Big],
\end{equation}
for integer $\alpha \ge 1$.  Here, $\rho_A$ is the reduced density matrix, $\rho_A = {\rm Tr}_B \{\rho\}$, and $A$ is a subregion of 
a lattice system (with $B$ being its complement). The von Neumann entanglement entropy corresponds to the limit $\alpha \to 1$.
 
The direct measure of Renyi entropies can not be done in QMC using conventional estimators, however recent work has demonstrated that it is possible to measure 
$S_{\alpha}$ for $\alpha \ge 2$ using a {\it replica trick} \cite{Holz,Cardy,Fradkin,BP,Naka}.  Here, $\alpha$ copies of the simulation cell are used, with modified 
periodic boundaries in the ``propagation'' (or imaginary time) direction.  The evaluation of Renyi entropies proceeds in formally different ways for $T=0$ and $T>0$ 
QMC methods \cite{Hastings10,Melko10}.  Namely, for $T=0$, the calculation of $S_{\alpha}$ proceeds in analogy with Eq.~(\ref{za2}) (where $\Psi$ is the wavefunction 
of the $\alpha$-times replicated system), with the operator $A$ replaced by a ``swap'' operator for $\alpha=2$ \cite{Hastings10} or permutation operator for 
$\alpha \ge 3$ \cite{Kallin11}. 

\begin{figure}
\includegraphics[width=7.0cm, clip]{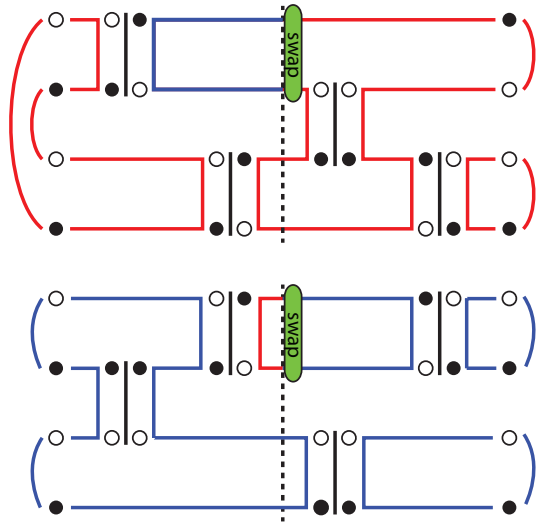}
\caption{The {\rm SWAP}$_A$ operator for calculating the Renyi entropy $S_2$, where subsystem $A$ consists of the two top sites. The simulated system is 
composed of two non-interacting replicas, top (compare to Fig.~\ref{loops}), and bottom, which are sampled independently. Just to the right of the mid-point 
of the configuration (dashed line) the {\rm SWAP}$_A$ operator (green vertical bar) connects the A spins in the top replica to the bottom one, as indicated by 
the colors of the loops \cite{Kallin11}.}
\label{swap}
\end{figure}

The actual simulation procedures are easiest to understand in the context of the projector algorithm discussed in Sec.~\ref{su2method}.  Here, the second Renyi 
entropy (for example) is given by the  mid-point evaluation of a SWAP$_A$ operator on two replicas of the system (see Fig.~\ref{swap}), that literally swaps states 
(spins or valence-bond endpoints) between replicas when they lie in region $A$.  Then,
\begin{equation}
 S_2 = -\ln \Big[ \langle {\rm SWAP}_A (\mathcal{L})  \rangle \Big],
\end{equation}
where the expectation value is calculated at the mid-point of the configuration, and expressed using the loops of the transition graph  $\mathcal{L}$ formed when 
superimposing the projected bra and ket states; Eq.~(\ref{vboverlap}).  Namely,
\begin{equation}
\frac{\langle V_l| {\rm SWAP}_A | V_r\rangle}{\langle V_l| V_r\rangle} = 2^{N_{\rm swap} - N_\circ } ,
\end{equation} 
where $N_{\rm swap}$ is the number of transition-graph loops of the swapped configuration, while $N_\circ$ is for the configuration before the swap 
operator is applied (Fig.~\ref{swap}). We refer to Ref.~\cite{Kallin11} for further details of this particular $T=0$ projector QMC implementation.  

At finite temperature, the explicit evaluation 
of the SWAP$_A$ operator can be replaced by an evaluation of the difference in free energies between a replicated system (similar to Fig.~\ref{swap}), 
and an un-replicated system \cite{Melko10}, using Eq.~(\ref{za1}).  Although this technique looks quite different formally, both the $T=0$ and $T>0$ techniques 
can be viewed on a similar footing when considered in the context of SW loop or cluster algorithms, much like the unified framework of the QMC itself, discussed
above in Sec.~\ref{ss:method}. In some cases, the implementation for these two different techniques can be made to be practically identical.  

\section{Ground states and quantum phase transitions in SU(2) models}
\label{sec:su2models}

It has been a matter of debate for a long time whether the non-linear $\sigma$-model can correctly capture a phase transition from the N\'eel state 
to a non-magnetic state in two dimensions for $S=1/2$ spins \cite{Chakravarty89}. Other terms, related to Berry phases of the spins, have to be included in the long wavelength description to obtain a VBS state \cite{Read90,murthy1990:mono}. It was recently proposed that the N\'eel and VBS states are separated
by a ``deconfined'' quantum-critical (DQC) point \cite{Senthil04a} described by the {\it non-compact} CP$^1$ field theory~\cite{Motrunich04} (in contrast to the nonlinear 
$\sigma$-model, which is equivalent to the {\it compact} CP$^1$ description, i.e., in the deconfined scenario the Berry phases render the gauge field effectively
non-compact at the critical point).
This field theory describes two flavors of spinons interacting with a non-compact U($1$) gauge field.  The VBS and N\'eel order parameter form due to confinement and condensation, 
respectively, of spinons, and therefore cannot be regarded as two separate order parameters at the phase transition.

In this section we first discuss the rather well-studied continuous quantum phase transition between the N\'eel state and a quantum paramagnet in systems 
with bimodal Heisenberg couplings forming a static pattern. In this case the paramagnet does not break any symmetries of the Hamiltonian and the transition 
is believed to fall in the 3D O($3$) universality class. A VBS state with spontaneously broken lattice symmetries can be achieved with frustrated 
interactions, which, however, are sign problematic in QMC simulations. The J-Q model \cite{Sandvik07} is a designer Hamiltonian circumventing this 
problem. We will review some of the recent work aimed at characterizing its VBS state and  N\'eel--VBS transition.

\subsection{N\'eel--paramagnetic transition in dimerized systems}
\label{sec:dimermodels}

Consider $S=1/2$ spins on a 2D bipartite lattice (e.g., the simple square lattice), interacting with antiferromagnetic Heisenberg exchange of 
strength $J_1$, written using the singlet projectors (\ref{psinglet}),
\begin{equation}
H_1 = -J_1 \sum_{\langle i,j\rangle} P_{ij}.
\label{h1def}
\end{equation}
Here $\langle i,j\rangle$ denotes nearest-neighbor spins. This system has long-range N\'eel order at $T=0$. Now introduce {\it dimerization}, 
by considering pairs (dimers) of spins $[i,j]$ such that each spin belongs exactly to one dimer. In addition to the nearest-neighbor couplings (\ref{h1def}),
let there be additional intra-dimer couplings of strength $\lambda$ (i.e., the total intra-dimer coupling is $J_1+\lambda$), such that the Hamiltonian is
\begin{equation}
H = H_1  -\lambda \sum_{[ i,j]} P_{ij} .
\end{equation}
If $\lambda>0$ and $J_1\to 0$, the ground state is clearly a product of singlets on the dimers---a quantum paramagnetic state that we will refer
to as a valence-bond-liquid (VBL), motivated by the fact that the state breaks no symmetries and is strongly interacting away from the large-$\lambda$
limit. When $\lambda=0$  the system is the standard Heisenberg model (\ref{h1def}). There is, thus, a quantum phase transition between N\'eel and VBL ground 
states as a function of $g=(J_1+\lambda)/J_1$. The dimers can be arranged in different ways, and instead of dimers one can also use larger $J_1$-coupled 
units of an even number of spins, e.g., $4$-site plaquettes, or one can use a bilayer, with $J_1$ and $J_2$ the inter-and intra-layer couplings, respectively.

\subsubsection{$T=0$ criticality}
Based on symmetry arguments alone, the N\'eel--VBL transition in these dimerized (or polymerized) systems should be in the classical 3D O($3$) 
universality class. Many QMC studies 
have been devoted to checking the exponents against available results for the classical transition \cite{Sandvik94,Troyer96,Matsumoto01,Wang06,Wenzel08,Wenzel09}.
For some of the systems, e.g., bilayers \cite{Wang06} and columnar dimers on the square lattice \cite{Matsumoto01,Wenzel09,Sandvik10b}, results have been 
obtained that rival classical calculations in precision (statistical error bars) and in general the exponents agree very well with each other. In other 
cases, e.g., with the dimers arranged in a staggered pattern, some studies initially indicated a new universality class \cite{Wenzel08}, while others
supported the O($3$) transition \cite{Jiang12}.  Later it was realized that these cases, where the dimer pattern lacks inversion symmetry, are associated 
with certain operators (cubic interactions) in the field-theory formulation  which are not present in systems with inversion symmetry \cite{Fritz11}. 
These operators are asymptotically irrelevant in the RG sense, but can lead to substantial finite-size corrections that can easily be mistaken for a 
different universality class.

\subsubsection{$T>0$ scaling}
Field-theoretic descriptions using the nonlinear $\sigma$-model \cite{Haldane83,Chakravarty89,Chubukov94} have given a wealth of 
predictions also for the $T>0$ quantum-critical ``fan'' extending out into the plane $(g,T)$  from the critical point $(g_c,0)$. Scaling 
behavior in this extended $T>0$ region is an important generic characteristic of quantum phase transitions, in contrast to classical phase transitions 
with typically very narrow regions of criticality. This aspect of dimerized models has also been investigated in detail in QMC 
studies \cite{Sandvik95,Brenig06,Sandvik11a}

\subsubsection{Experimental realizations}
A N\'eel--VBL transition of the type discussed above can in principle be experimentally realized in systems consisting of coupled dimers as a function 
of pressure. The authors are not aware of any material where the phase transition can be crossed in a quasi-2D systems of weakly 
coupled planes, but a well-studied 3D case is TlCuCl$_3$ \cite{Cavadini01,Ruegg04}. Here neutron scattering shows a transition from a paramagnet 
to a N\'eel state at a critical pressure \cite{Ruegg08}, including the analogue of the Higgs boson (longitudinal mode) in the ordered 
state \cite{Sachdev09}. In this case the values of the couplings and their pressure dependence are not known in detail, but one can still learn a lot from 
studies of designer Hamiltonians by focusing on universal properties \cite{Troyer97,Yasuda05,Yao07,Jin12,Oitmaa11}.

\subsection{Deconfined N\'eel--VBS transition in J-Q models}
\label{ss:jq2}
While spontaneously formed VBS states in translationally invariant 2D spin systems have been discussed for more than two decades \cite{Chandra88,Dagotto89,Read89},
most of the early work focused on frustrated systems \cite{Dagotto89,Schulz96,Capriotti01}, for which it is very difficult to carry out unbiased numerical 
calculations. The proposal of a DQC point separating the N\'eel and VBS states motivated detailed numerical studies of this transition, and, thus, construction 
of suitable designer Hamiltonians amenable to QMC simulations. Numerical work preceding the theory \cite{Sandvik02} had already indicated the intriguing possibility 
of continuous magnetic--VBS quantum phase transitions (where standard arguments based on Landau-Ginzburg theory predict generically first-order transitions) in
U($1$) (quantum XY) models. Tractable SU($2$) spin models exhibiting N\'eel--VBS transitions were lacking, however, until the introduction of the J-Q class of 
models \cite{Sandvik07}, where $J$ refers to Heisenberg exchange on a bipartite lattice and $Q$ is a multi-spin interaction which competes against N\'eel 
order but does not generate a sign problem. 

\begin{figure}
\includegraphics[width=6cm, clip]{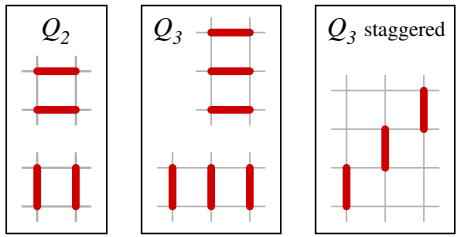}
\caption{$Q_2$ and $Q_3$ interactions on the square lattice. The bars indicate the locations of singlet projectors 
$C_{ij}$ on site pairs $ij$. The $Q_2$ and $Q_3$ arrangements in the left and center panels lead to columnar VBS states,
while the right $Q_3$ arrangement leads to a staggered VBS. The Hamiltonian contains all unique translations and
$90^0$ rotations (and in the staggered case also reflections) of the operator patterns.}
\label{qterms}
\end{figure}

The $Q$ interactions are most conveniently expressed in terms of the singlet projectors, Eq.~(\ref{psinglet}), in the simplest case on the 2D square lattice;
\begin{equation}
H_Q = - Q\sum_{\langle ijkl\rangle } P_{ij}P_{kl},
\label{hqdef}
\end{equation}
where the site pairs $ij$ and $kl$ form parallel edges (with both horizontal and vertical arrangements included) on a $2\times 2$ plaquette. 
One can also consider more than two projectors in the product and denote by $Q_p$ an interaction with $p$ projectors. Fig.~\ref{qterms} illustrates $Q_2$ 
and $Q_3$ terms leading to columnar VBS ground states. In addition to the number of projectors in the product, they way they are arranged relative to each 
other on the lattice is also crucial, with different patterns leading to different VBS ground states of the ``pure $Q$'' models (\ref{hqdef}). VBS to N\'eel 
transitions can be studied in $J$-$Q$ models, $H=H_J+H_Q$, as a function of the ratio $J/Q$ of the standard Heisenberg exchange and the multi-spin 
interaction.

A staggered $Q_3$ term (illustrated in Fig.~\ref{qterms})
on the square lattice was investigated in Ref.~\cite{Sen10}. This leads to a strongly first-order transition. In contrast, 
the columnar $Q_2$ and $Q_3$ arrangements in Fig.~\ref{qterms} lead to continuous transitions (or, in principle, the transition could be very weakly first-order). 
Various types of $Q$ terms have also been investigated in the honeycomb lattice \cite{Banerjee11}, with similar results. In analogy with quantum 
dimer models \cite{Rokhsar88,Moessner01}, first-order transitions correspond to VBS states that do not support any local dimer fluctuations, while 
the continuous transitions are into VBSs where such fluctuations exist. Alternatively, one can relate the different behaviors to different types of 
topological defects in the VBS \cite{levin2004:vbs,Banerjee11}.

\subsubsection{$T=0$ Critical behavior}

\begin{figure}
\includegraphics[width=8cm, clip]{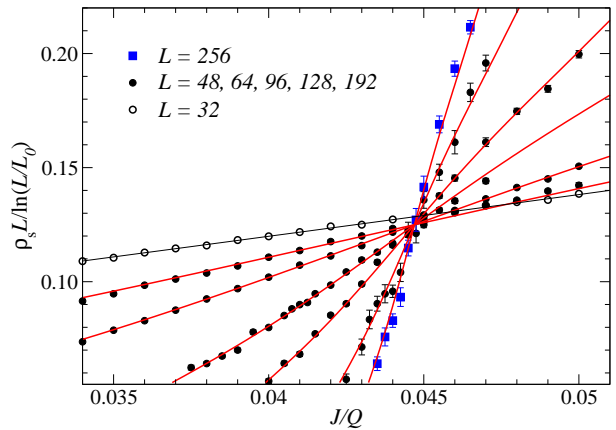}
\caption{Finite-size scaling of the spin stiffness of the $J$-$Q_2$ model in the vicinity of its quantum-critical point; from Ref.~\cite{Sandvik10c}. The 
calculations were done at inverse temperature $\beta=L$. The standard critical scaling form $L\rho_s \sim {\rm constant}$ has been modified by a log-correction 
(with $L_0=0.9$) in order to compensate for a weak drift of the crossing points. The red curves, which cross at a single point, the critical point $q_c=(J/Q)_c = 0.0447(2)$, 
represent a common scaling function for $L\ge 48$ (approximated by a common polynomial fitted to all the data points), $f[(q-q_c)L^{1/\nu}]$, with $\nu=0.59$.}
\label{jqrhos}
\end{figure}

The critical behavior of both the $J$-$Q_2$ and $J$-$Q_3$ models (with the columnar dimer arrangements in Fig.~\ref{qterms}) on the simple 2D square 
lattice has been analyzed in detail in several different ways \cite{Sandvik07,Jiang08,kaul2008:jqlgN,melko2008:jq,lou2009:sun,Sandvik10c}. One useful quantity 
for studying the destruction of the N\'eel state is the spin stiffness $\rho_s$. It should scale at a critical point as $\rho_s \sim L^{2-d-z}$, 
where $z$ is the dynamic exponent. The DQC theory is Lorenz-invariant, i.e., $z=1$. One can test this prediction by graphing 
$L\rho_s(L)$ versus the coupling ratio $J/Q$ for different $L\times L$ lattices (in the ground state or with the inverse temperature $\beta \propto L^z$). 
Such curves should cross at the critical point. In practice, crossing points often exhibit some drift with $L$ \cite{melko2008:jq}, reflecting scaling 
corrections \cite{Sandvik10c}. Results for the $J$-$Q_2$ model are graphed in Fig.~\ref{jqrhos} and discussed next.

In the $J$-$Q$ models the $L\rho_s$-crossing drift is anomalously large, which in one study \cite{Jiang08} had been interpreted as a weakly first-order transition (for which 
$L\rho_s$ should eventually diverge as $L$, due to coexistence of a stiff and a non-stiff phase). Another possibility is that the model has unusually
large scaling corrections, that could be intimately associated with the nature of the DQC point. For example, in the DQC theory the perturbation to the 
critical point leading to VBS order (a quadrupled monopole operator) is dangerously irrelevant. It could potentially lead to large, but ultimately conventional
scaling corrections. Another possibility is that the corrections are logarithmic (multiplicative) \cite{Sandvik10c,banerjee2010:log}, similar to those known 
in the critical Heisenberg chain \cite{Giamarchi89}. Such corrections do not appear in the large-$N$ calculations of the CP$^{N-1}$ theory, but potentially they could 
appear for small $N$, as has recently also been argued for based on a modified version of the DQC theory \cite{Nogueira11}. Numerical data for the $J$-$Q_2$ 
model \cite{Sandvik10c}, on $L\times L$ lattices with $L$ up to $256$, can be well accounted for by a log-correction, as shown in Fig.~\ref{jqrhos}, but the 
deviations from pure scaling can also be fitted with a conventional multiplicative correction of the form $(1+aL^{-\omega})$ with large $a$ and small $\omega$.

\subsubsection{$T>0$ critical scaling} 

The quantum-critical ``fan'' in the $(J/Q,T)$ plane has also been investigated. Initial calculations confirmed that
the scaling is governed by a $z=1$ critical point \cite{melko2008:jq}. Later studies \cite{Sandvik10c,Sandvik11a} noted log-like corrections also here, e.g., 
while the conventional $z=1$ scaling implies that the uniform magnetic susceptibility scales linearly in $T$ \cite{Chakravarty89}, the behavior actually appears 
to be of the form $\chi \sim T[1+a\ln(1/T)]$. These corrections, like those at $T=0$ discussed above, remain an intriguing unexplained aspect of the N\'eel--VBS 
transition in the $J$-$Q$ models.

\subsubsection{Emergent U($1$) gauge fluctuations}

An important and remarkable aspect of the DQC theory is the emergent conservation law of the gauge flux, due to the irrelevance of monopoles at the critical point. 
There is a U($1$) symmetry associated with this conservation law at the critical point, which is also emergent and not explicitly contained in 
the $J$-$Q$ model, which only has an SU($2$) spin-rotational symmetry, the discrete symmetries of the square lattice and time reversal. The emergent U($1$) 
symmetry can, however, be observed in the fluctuations of the VBS order parameter \cite{Sandvik07}. Naively one would expect these fluctuations 
to reflect the Z$_4$ symmetry of the VBS (which is dictated by the square lattice), but instead the symmetry group expands to U($1$) upon approach to the 
critical point. The interpretation of this emergent U($1$) VBS symmetry is that it reflects the non-compactness of the gauge field (``photon'') 
of the DQC theory. Here we will review how this emergence is detected in QMC simulations.

Consider the VBS (dimer-dimer) correlation function,
\begin{equation}
C_{\alpha} ({\bf r}_{ij}) = \bigl \langle B_\alpha({\bf r}_i)B_\alpha({\bf r}_j) \bigr \rangle, 
\label{cddef}
\end{equation}
where ${\bf r}_{ij}={\bf r}_{i}-{\bf r}_{j}$ is the spatial separation of the operators $B_\alpha$, $\alpha=\hat x,\hat y$,
which measure the spin correlations on nearest-neighbor bonds in the $\alpha$ direction;
\begin{equation}
B_{\hat x}({\bf r})={\bf S}({\bf r}) \cdot {\bf S}({\bf r}+\hat {\bf x}),~~~~~~B_{\hat y}({\bf r})={\bf S}({\bf r}) \cdot {\bf S}({\bf r}+\hat {\bf y}).
\label{bxdef}
\end{equation}
We define the expectation value of this operator when averaged over all sites ${\bf r}=(x,y)$; $b_\alpha=\langle B_{\alpha}(x,y)\rangle \not = 0$. In a VBS, the 
expectation value acquires a spatial modulation. With the dimers (stronger bonds) oriented in the $x$ direction, the asymptotic form of the spatially averaged 
$x$ dimer correlations in a columnar VBS is $C_{\hat x}(x,y) = b_{\hat x}^2+(-1)^{x}(\hbox{$\frac{1}{2}$}D^2+a{\rm e}^{-r/\xi})$, where $r=(x^2+y^2)$, $D$ is the 
magnitude of the order parameter, and $a$ is a constant. The rate of decay of the oscillating term toward a non-zero value as $r\to \infty$ defines the correlation 
length $\xi$ (where we suppress the directional dependence of $\xi$, which can be parameterized by $\xi_x$ and $\xi_y$). This is the standard definition of a 
correlation length, as the length-scale associated with the order parameter. Normally this (the larger of $\xi_x,\xi_y$) is the largest length-scale in the system. 

In a symmetry-broken $x$-oriented VBS, we can consider also dimer correlations in the direction $y$ perpendicular to the dimers. The asymptotic form of these 
correlations is $C_{\hat y}(x,y) = b_{\hat y}^2 + k{\rm e}^{-r/\Lambda}$, where $k$ is another constant (and again there are really two decay constants; $\Lambda_x$ and
$\Lambda_y$). These correlations are, at first sight, not associated with the order parameter, and the length-scale $\Lambda$ would then normally not be referred 
to as ``the correlation length''. However, $\Lambda$ can actually be larger than $\xi$, and this is at the heart of the emergent U($1$) symmetry of the VBS. 
According to the DQC theory, both lengths diverge as the critical point is approached, and $\Lambda \sim \xi^{1+a}$ with $a>0$. Close to the critical point, 
where the order parameter $D$ is small and $\Lambda \gg \xi$, on length scales less than $\Lambda$ the system will appear to have both $x$ and $y$ VBS order.

One can also associate $\Lambda$ with fluctuations of the {\it angle} of the order parameter. Since the VBS order parameter is a vector, ${\bf D}= (D_x,D_y)$, 
one can define a magnitude $D$ and an angle $\Theta$. $\Lambda$ defines a length scale below which the order parameter exhibits U($1$) symmetry---the system
not only has both $x$ and $y$ order up to this length scale, but one can think of the angle as fluctuating uniformly, even though it has to take one of the 
values $n\pi/2$ when the order parameters $D_x$ and $D_y$ of a symmetry-broken state are averaged over regions of size $\gg \Lambda$. There are many interesting 
consequences of the emergent U($1$) symmetry of the VBS order parameter, as further discussed in Ref.~\cite{Sandvik12}.

This phenomenon of an emergent symmetry higher than that of the order parameter in an ordered state is in fact a general aspect of systems with 
dangerously irrelevant perturbations. These perturbations reduce the symmetry of a system in such a way that the standard critical exponents 
at a phase transition into an ordered state are unaffected, but the symmetry broken in the ordered state is reduced \cite{Jose77,Oshikawa00}. A 
prototypical example is the 3D classical XY model with an added on-site potential $h\cos(q\theta_i)$ on all sites $i$, where $\theta_i$ is the spin 
angle and $q$ an integer \cite{Carmona00}. This model orders at a critical temperature $T_c(h)$, but, regardless of the value of $h$, the critical exponents 
are fixed at those of the standard XY transition. However, instead of breaking the continuous U($1$) symmetry, as the original $h=0$ model does, 
now only a $q$-fold (Z$_q$) symmetry is broken. As in the VBS discussed above, there is a length-scale $\Lambda \sim \xi^{1+a(q)}$ in the ordered state, which 
is associated with U($1$) symmetric fluctuations of a course-grained order parameter (magnetization) ${\bf m}=(m_x,m_y)$. Finite-size scaling procedures have been 
developed \cite{Lou07} to access the exponent $a(q)$ directly based on the cross-over from U($1$) to Z$_q$ symmetry seen in the order-parameter distribution 
$P(m_x,m_y)$. Such techniques have also been employed to study the VBS fluctuations of the $J$-$Q$ model, as we discuss next.

\begin{figure}
\includegraphics[width=7.5cm, clip]{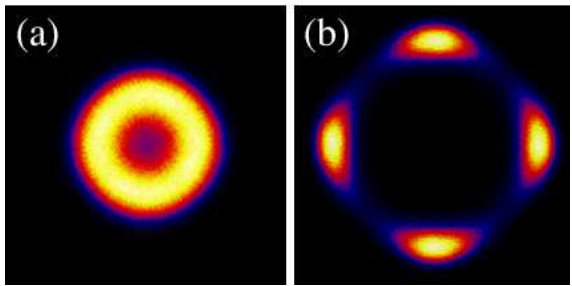}
\caption{Probability distribution $P(D_x,D_y)$ of the VBS order parameter of the $J$-$Q_3$ model on a $32\times 32$
lattice at $J/Q_3 = 0.575$ in (a) and $0.1765$ in (b). The critical ratio is $(J/Q_3)_c \approx 0.67$. Brighter colors
correspond to higher probability. Adapted from Ref.~\cite{lou2009:sun}.}  
\label{jq3histo}
\end{figure}

In simulations of a VBS on a finite lattice, where the Z$_4$ symmetry is not broken, one can study the emergent U($1$) and the length-scale $\Lambda$ by 
accumulating the probability distribution $P(D_x,D_y)$, where $D_x$ and $D_y$ are defined on the whole system in terms of matrix elements of the operators 
$B_\alpha$ in Eq.~(\ref{bxdef}), evaluated for individual configurations using transition-graph loops, as explained in Sec.~\ref{sec:su2models}, and Fourier 
transforming at $(\pi,0)$ and $(0,\pi)$ for $\alpha=\hat x$ and $\alpha=\hat y$ correlations, respectively. Fig.~\ref{jq3histo} shows results for the $J$-$Q_3$ 
model at two different ratios $J/Q_3$ \cite{lou2009:sun}. Far away from the critical point the histogram exhibits a four-fold symmetry consistent with a columnar 
VBS, while closer to the transition point the distribution is ring-shaped. The radius of the ring corresponds to the magnitude $D$ of the order parameter, and the 
absence of four-fold symmetry implies that here the system size $L=32 < \Lambda$. By analyzing the U($1$)--Z$_4$ cross-over as $L$ increases, using finite-size 
scaling techniques, one can extract the exponent governing the length-scale $\Lambda$. In Ref.~\cite{lou2009:sun} this resulted in  $\Lambda \sim \xi^{1+a}$, with 
$a \approx 0.2$

\subsubsection{Spinon deconfinement}

The emergent U($1$) length scale discussed above should also be reflected in the confinement of spinons (e.g., the size of a bound state) in the VBS state. By utilizing 
properties of the valence-bond basis, discussed in Sec.~\ref{su2method}, and generalizations to non-zero total magnetization \cite{Banerjee10b,Wang10}, it is 
possible to directly access $S=1$ excitations of VBS states and study their structure in terms of bound or unbound (deconfined) spinons. Such calculations have 
already been used to characterize deconfined spinons using a designer Hamiltonian with a VBS state in one dimension \cite{Tang11a}. 
Similar calculations for 2D J-Q models should shed further light on the nature of the spinons in the vicinity of the DQC point.

Another approach to confirming spinon excitations, which has already been employed for the 2D J-Q model, is to study the thermodynamic properties at 
criticality and compare them with predictions for a simple model of a gas of thermally excited bosonic spinons \cite{Sandvik11a}. The results show a remarkable 
consistency among different quantities which are related through expressions for the spinon gas, thus lending strong support to the deconfinement scenario.

\section{N\'eel-VBS transition in SU($N$) spin models}
\label{sec:sunmodels}

In this section we will discuss quantum criticality in SU($N$) spin models on 2D square lattices, as systematic generalizations of the SU($2$) systems 
discussed above. Secs.~\ref{ss:j1N}, \ref{ss:jqN}, \ref{ss:j1j2N}, and \ref{ss:j1j2q} describe several studies carried out on a single layer, and Sec.~\ref{ss:bilN} summarizes 
a recent study on a bilayer. 

\subsection{SU($N$) Heisenberg models}
\label{ss:j1N}
A natural extension of the SU($2$) Heisenberg model is to spin models with SU($N$) symmetry. Such generalizations were initially introduced to
facilitate access to a solvable large-$N$ limit~\cite{affleck1985:lgN,Read89}. In the meantime, new physical systems have also motivated an interest 
in such models at arbitrary finite values of $N$~\cite{gorshkov2010:sun,kugel1982:kk}, leading to extensive studies of SU($N$) spin systems. The 
physical degrees of freedom of a particular spin model depend not only on the global symmetry but also on the chosen representation under which the 
spins transform. Thus, there are a number of distinct extensions of the SU($2$) model to SU($N$), depending on what properties are intended to be retained 
in the generalization. 

In the context of the study of the DQC point at the N\'eel-VBS transition, the ability of two spins to form a singlet is the most
central characteristic required to form a VBS. It is possible to choose SU($N$) models which have this property by working on bipartite lattices and
assigning SU($N$) spins that transform as relative conjugate representations on each of the sublattices, as discussed in Sec.~\ref{ss:su2N}. In such 
models two spins on opposite sublattices can form a singlet, allowing for the possibility of a valence-bond solid.  The simplest 
SU($N$) invariant two-site sign-problem free Hamiltonian interaction one can define for such models is given by $H_1$ in Eq.~(\ref{h1def}), with $P_{ij}$ 
generalized to the SU($N$) singlet projector defined in Sec.~\ref{ss:su2N}. We note again that for $N=2$ this reduces to the familiar $S=1/2$ 
antiferromagnetic Heisenberg model. 

The Hamiltonian Eq.~(\ref{h1def}) has been studied extensively as a function of $N$. On a 1D chain the well known critical Bethe state for $N=2$ gives way to a 
VBS ordered state for all $N\geq 3$~\cite{barber1989:d1n3_vbs,klumper1989:d1n3_vbs,affleck1985:lgN}. The 2D version of $H_{1}$ was shown 
in the large-$N$ limit to map onto a quantum dimer model with only a kinetic term \cite{read1989:nucphysB}. From direct numerical studies of the quantum dimer model, 
it is known that the kinetic-only model orders into a VBS \cite{sachdev1989:qd_vbs, Syljuasen06}.  This implies that in $H_{1}$ there must be a transition between the 
magnetic state at $N=2$ and the VBS at $N=\infty$.  The location of the transition was first determined to lie between $N=4$ and $N=5$ 
by QMC simulations~\cite{harada2003:sun,Kawashima07}.

\subsection{SU($N$) $J_1$-$Q$ model for $N\leq 4$}
\label{ss:jqN}

In the previous section we saw that on the square lattice the N\'eel-order found for $N=2,3,4$ gives way to VBS
order for $N\geq 5$ in the SU($N$) antiferromagnet, Eq~(\ref{h1def}). In order to access the quantum phase transition by QMC simulations at a 
given value of $N\leq 4$ one needs new sign-problem free ``designer'' couplings that destroy the N\'eel phase and result in the VBS. The first such term discovered 
was the $Q$-interaction, Eq.~(\ref{hqdef}), introduced for SU($2$) in Sec.~\ref{ss:jq2}. It can be generalized for any $N$, with $P_{ij}$ again 
the SU($N$) singlet projectors. Since the matrix elements of $P_{ij}$ are all positive, such a $Q$ coupling is free of the QMC sign problem. 

It turns out that the $Q$-only model, with the projectors arranged as in Fig.~\ref{qterms}, is always VBS ordered, so that for the cases of $N=2,3,4$, 
the $J_1$-$Q$ model has a N\'eel-VBS transition~\cite{Sandvik07,lou2009:sun}. Detailed QMC studies for these $N$ find strong evidence for a 
continuous transition between the N\'eel and VBS ground states \cite{melko2008:jq,kaul2011:su34,banerjee2010:log,banerjee2010:su3}. 
As discussed above for the SU($2$) case, certain observables show deviations from standard scaling laws for the system sizes studied, that could either 
be due to conventional corrections to scaling or actual violations of the scaling laws (but they do not resemble violations of the kind that would be 
expected at a first-order transition \cite{Sandvik10c}). The origin of these corrections remains to be resolved.

\subsection{SU($N$) $J_1$-$J_2$ model for $N\geq 5$}
\label{ss:j1j2N}

For $N\geq 5$, the $Q$ term strengthens the VBS that is already present in the $J_1$ only model, and hence to study the N\'eel-VBS transition 
for large-$N$, a new designer Hamiltonian is needed. To supply such a model, an interaction $J_2$ can be included between sites on the same sublattice 
which stabilizes the N\'eel state \cite{kaul2011:j1j2};
\begin{equation}
H_{2}= -J_2 \sum_{\langle\langle ij\rangle\rangle} \Pi_{ij},
\end{equation}
where the $\Pi_{ij}$ interaction was introduced in Eqs.~(\ref{pija}) and (\ref{pijb}).
In order to see that this interaction favors the N\'eel state, we note that in the $H_{2}$ model by itself, the two sublattices 
are decoupled. The ground state corresponds to having an independent SU($N$) ferromagnet on each sublattice. Turning on a small $J_1\ll J_2$ 
interaction will clearly cause the sublattice degeneracy to be lifted, resulting in the two independent ferromagnets to lock into a single
antiferromagnetic state. These arguments are true independent of the value of $N$. Thus, for every $N\geq 5$ the VBS state is realized when 
$J_1\gg J_2$ and the N\'eel state is obtained when $J_2 \gg J_1$. Thus, there must be at least one quantum phase transition as the ratio
$J_2/J_1$ is tuned. From numerical simulations of the $J_1$-$J_2$ model, there is compelling evidence for a direct transition between the two 
phases, with no indication of an intervening phase \cite{kaul2011:j1j2}. 

\subsection{$J_1$-$J_2$-$Q$ model; anomalous dimensions}
\label{ss:j1j2q}

\begin{figure}
\includegraphics[width=9cm, clip]{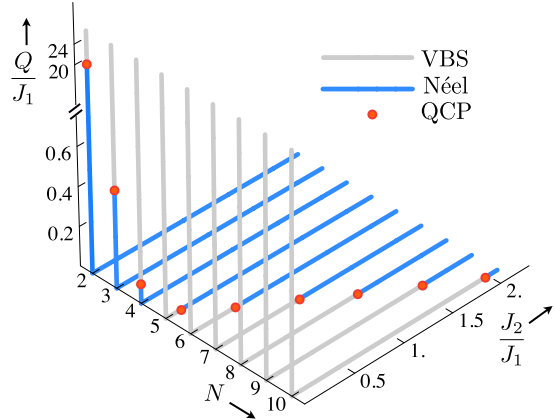}
  \caption{ \label{fig:pdj1j2q} Phase Diagram of the SU($N$) symmetric sign-problem free
    $J_1$-$J_2$-$Q$ model as a function of $N$. Each of the couplings has been introduced
    in the text: the generalized nearest-neighbor antiferromagnetic exchange $J_1$ in Sec.~\ref{ss:j1N}, the four-spin coupling $Q=Q_2$ in 
    Sec.~\ref{ss:jqN} and Fig.~\ref{qterms}, and the generalized ferromagnetic next-nearest-neighbor coupling $J_2$ in Sec.~\ref{ss:j1j2N}. 
    This SU($N$) antiferromagnet allows for an unbiased study of deconfined quantum
    criticality at the N\'eel-VBS transition for each value of $N$. }
\end{figure}

One can also combine the $J_1$-$J_2$ and $J$-$Q$ model. The phase diagram of the resulting $J_1$-$J_2$-$Q$ model as a functions of $N$ is
summarized in Fig.~\ref{fig:pdj1j2q}. So far, the phase boundaries have only been computed in the two perpendicular planes shown in the figures.
We here discuss some aspects of the critical behavior for $N \in \{2,3,\ldots,12\}$ obtained in these calculations and relate the results to
large-$N$ expansions within the DQC theory.

An important prediction of the DQC scenario~\cite{Senthil04a} at the N\'eel-VBS transition is that both order parameters are
simultaneously quantum critical at the phase transition and that space and time scale in the same way, i.e., the dynamic critical 
exponent $z=1$. This implies that both correlation functions decay as Lorentz-invariant power laws,
\begin{equation}
\label{eq:expdef}
C_{N,V} ({\bf r},\tau) \sim  \frac{1}{~(r^2+c^2\tau^2)^{(1+\eta_{V,N})/2}}.
\end{equation}
 where $C_N$ and $C_V$ are the two point correlation functions of the
 N\'eel and VBS order parameters. The indices $\eta_N$ and $\eta_V$
 are the so-called anomalous dimensions of the N\'eel and VBS order
 parameters. The DQC scenario predicts that
 the continuum field-theoretic universality of the N\'eel-VBS critical
 point is described by the non-compact CP$^{N-1}$ universality. Quantitative
 estimates for the universal indices that characterize this
 universality class are available only in the large-$N$
 limit~\cite{halperin1974:largeN}. The ${1}/{N}$ expansions for $\eta_N$~\cite{kaul2008:u1} and $\eta_V$~\cite{murthy1990:mono,metlitski2008:mono} are,
\begin{equation}
\label{eq:oneonN}
\eta_N = 1 - \frac{32}{\pi^2N},~~~~~~
1+\eta_V = 2 \delta_1 N~~~(\delta_1\approx 0.24).
\end{equation}
We note here that as $N\rightarrow\infty$, $\eta_N \rightarrow 1$ and $\eta_V\rightarrow \infty$. Both results are very unusual, since typically 
$\eta$-exponents are small. For reference, in the O($N$) model $\eta\rightarrow0$ in the $N\rightarrow\infty$ limit and $\eta \approx 0.037$ 
for the O($N=3$) model~\cite{campostrini2002:o3}.

\begin{figure}
\includegraphics[width=8.4cm, clip]{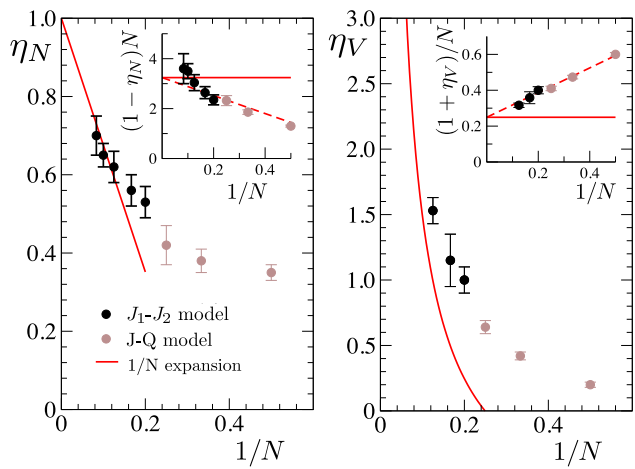}
 \caption{ \label{fig:exp} Anomalous dimensions of the N\'eel (left) and VBS (right)
  order parameters as a function of $N$. The main panels show $\eta_N$ and $\eta_V$ versus $1/N$. For $N=2,3$ and $4$, the data are 
  for the $J$-$Q$ model \cite{lou2009:sun}, and the results for $N>4$ are for the $J_1$-$J_2$ model \cite{kaul2011:j1j2}. The analytic 
  results from the $1/N$ expansion of the CP$^{N-1}$ field theory, Eq.~(\ref{eq:oneonN}), are shown as thick red lines. The left and 
  right insets show $N(1-\eta_N)$ and $(1+\eta_V)/N$, respectively. These quantities must be finite in the  $N\rightarrow \infty$ 
  limit according to the DQC theory and should be given by the prefactors of the $N$-dependent terms in Eq.~(\ref{eq:oneonN}). The values
  are indicated by the solid horizontal lines. The next corrections to the exponents have not been computed analytically yet, but we can estimate 
  them approximately as $\eta_N = 1+32/(\pi^2 N)-3.6(5)/N^2$ , $1+\eta_V = 0.2492 N + 0.68(4)$ (shown as dashed fitted lines in the insets). 
  Graphs reproduced from Ref.~\cite{kaul2011:j1j2}.}
\end{figure}

We now turn back to the study of the lattice designer $J_1$-$J_2$-$Q$ Hamiltonian. By analyzing the size dependence of the correlation 
functions in Eq.~(\ref{eq:expdef}) at the location of the critical points shown in Fig.~\ref{fig:pdj1j2q}, it is possible to
estimate values for $\eta_N$ and $\eta_V$ as a function of $N$~\cite{lou2009:sun,kaul2011:j1j2}. Results for $2\leq N \leq 12$ are 
summarized in Fig.~\ref{fig:exp}. The lattice Hamiltonians reproduce the ``smoking gun'' DQC feature that $\eta_N \to 1$ and $\eta_V \to \infty$ in 
the $N\rightarrow \infty$ limit. Furthermore, extrapolations of the data, with the constant terms subtracted off, show quantitative 
agreement, within a few percent, with the corrections in the large-$N$ expansions, Eq.~(\ref{eq:oneonN}). Note that the correct description for large 
$N$ is far from trivial, since the continuum theory was not derived from the microscopic Hamiltonians considered here but was posited as a 
generic theory of the N\'eel--VBS transition. 

The fact that the exponents agree so well lends strong positive support in favor of the DQC scenario for large $N$. Clearly, it would be 
very interesting to compute further $1/N$ corrections analytically, to test the theory in this quantitative way down to the smallest $N$. We note 
that the exponent $\eta_N$ for the SU($2$) case has been estimated based on classical Monte Carlo simulations of the non-compact CP$^1$ model, and
also of the 3D Heisenberg model with ``hedgehog'' topological defects suppressed (which should lead to the same universality) \cite{Motrunich04}. 
The exponents are in rough agreement with those of the SU($2$) J-Q model, but it would be very interesting to push these classical simulations to 
higher precision as well. Another study of a classical model argued to realize the DQC action was found to have a first-order 
transition \cite{Kuklov08}.

\begin{figure}
\includegraphics[width=8.4cm, clip]{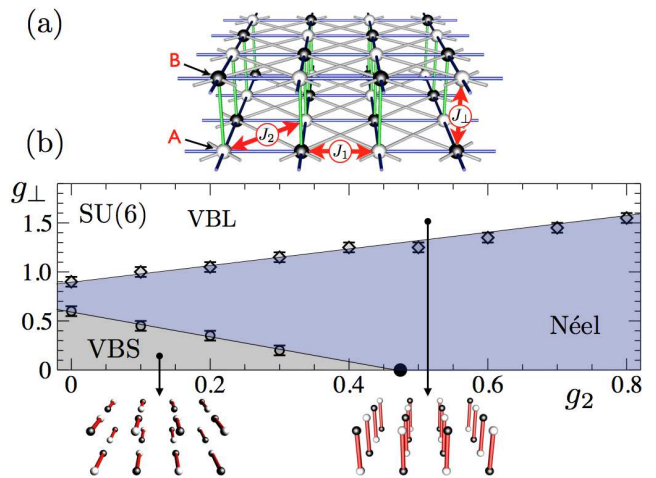}
  \caption{ \label{fig:pd_bil}  Physics of the SU($N$) bilayer model.  (a) Lattice and couplings. White sites form the A sublattice with 
  SU($N$) spins that transform in the fundamental representation. Black sites on the B sublattice transform as the conjugate to this representation. (b) Phase diagram of the
  SU($6$) bilayer model as a function of $g_\perp =J_\perp/J_1$ and $g_2=J_2/J_1$. The black circle at $g_\perp=0$ is the N\'eel-VBS quantum critical point. Remarkably, 
  in the bilayer geometry with $g_\perp\neq 0$ this continuous transition becomes first-order (indicated by open circles), due to the cancellation of Berry phases between 
  the layers. The VBL phase at large $g_\perp$ is a featureless paramagnet with singlets predominantly forming between the planes. Cartoons of the basic singlet physics 
  in the symmetry-broken VBS (left) and uniform VBL (right) phases are shown. The SU($N$) N\'eel-VBL transition is discussed at length in Ref.~\cite{kaul2012:sun_bil}.}
\end{figure}

\subsection{SU($N$) Bilayer}
\label{ss:bilN}

The study of the destruction of SU($N$) magnetic order in the bilayer geometry, illustrated in Fig.~\ref{fig:pd_bil}(a), provides an interesting 
example to test our understanding of the role of Berry phases at DQC points, since these should cancel between the two square lattice layers. 
Consider a bilayer in which each layer is described by some combination of the $J_1$-$J_2$ interactions discussed in the previous section and with
the following coupling between the layers,
\begin{equation}
 H_{J_\perp} = -J_\perp \sum_{[ij]} P_{ij},
\end{equation}
where the sum over $[ij]$ is taken between spins vertically next to each other in the bilayer (i.e., $[ij]$ form dimers, as in Sec.~\ref{sec:dimermodels}). For 
any $N$, the $J_\perp$-only model has a featureless non-degenerate ground state consisting of a product of vertical valence bonds (singlets)---the SU($N$)
generalization of the VBL state discussed in Sec.~\ref{sec:dimermodels}. The introduction of small intra-layer couplings like the $J_1$-$J_2$ interactions will 
make the ground state deviate from a simple product state by increasing the amount of entanglement, but these couplings cannot destabilize the VBL paramagnet,
as long as they are small compared to $J_\perp$; see Fig.~\ref{fig:pd_bil}(b). 

\subsubsection{N\'eel--VBL transition}

When the value of $J_2/J_1$ places the system within the N\'eel phase when $J_\perp=0$, there is a transition from the N\'eel state into the VBL for increasing 
$J_\perp$. This transition in the SU($2$) case is continuous and belongs to the 3D O($3$) universality class, as discussed in Sec.~\ref{sec:dimermodels}. For $N>3$, 
QMC calculations \cite{kaul2012:sun_bil} show that the transition is first-order, however, as also predicted by a mean-field theory. For $N=3$ the transition appears 
continuous based on the largest system sizes studied. If this is indeed the case, it would be in the universality class of the compact CP$^2$ model~\cite{Nahum11}. 
Further studies of the $N=3$ model should be carried out to resolve this issue.

\subsubsection{N\'eel--VBS transition}
A very interesting aspect of the SU($N$) bilayer models is the limit where $J_\perp$ is weak and there is a N\'eel--VBS transition as a function of $J_2/J_1$.
It is well known that the Berry phases cancel in the long-wavelength limit between the layers. Nevertheless, both N\'eel and VBS phases must clearly be stable 
to a small finite interlayer coupling. What is the fate of the $J_\perp=0$ deconfined N\'eel-VBS transition in the presence of the bilayer coupling? Remarkably, 
the QMC simulations show that the transition between the very same N\'eel and VBS phases becomes first-order in the bilayer geometry. This finding can be 
understood as a restoration of the Landau paradigm, due to the cancellation of Berry phases which is a relevant perturbation at the deconfined critical 
point \cite{kaul2012:sun_bil}. This is another piece of strong evidence for the correctness of the DQC theory, as well as the more general field-theoretical 
understanding of the role of Berry phases.

\section{Spin liquids and deconfinement in U($1$) models}
\label{sec:u1models}

One of the main foci of research into designer Hamiltonians is the search, detection and characterization of quantum spin liquid (QSL) phases.  That is, although there 
has been much progress in understanding properties of QSL states from effective field theories, there are very few microscopic models amenable to unbiased calculations 
that can be argued to exhibit a QSL without controversy. The recent spate of high-profile numerical studies identifying possible QSL candidate Hamiltonians \cite{Yan,Meng10,J1J2} 
has sparked intense activities on the theory side, highlighting the synergy between analytical theory and numerics. Thus, we are interested in finding the simplest models 
harboring QSL states, that are also amenable to large-scale simulation by QMC.

Much of the reason for the scarcity of microscopic models is that fundamental ingredients required to give rise to QSL phases, namely geometrical frustration, are also typical 
harbingers of the sign problem.  Despite this, the absence of the sign problem does not fundamentally {\it preclude} the existence of spin liquid physics.  This is most evident 
in Hamiltonians with U($1$) symmetry---anisotropic spin (or Bose-Hubbard) models specifically designed to capture the essential ingredients needed to promote a QSL phase without 
causing the sign-problem. A crucial fact in this regard is that there are no limitations on the diagonal couplings, because the sign problem is only caused by the off-diagonal
matrix elements in the chosen QMC basis (see Sec.~\ref{sec:methods}). Thus, diagonal frustration can be simulated in any form.

Examples of both 2D gapped \cite{Isakov1,Isakov2,Long,TopoEE} and 3D and gapless \cite{Isakov3} QSLs have recently been studied with large-scale QMC simulations of 
such sign-problem free U($1$) models. In this section, we will explore one class of Hamiltonians that contains the simplest type of QSL phase, which can be related to a 
$Z_2$ gauge theory. We will demonstrate that universal topological properties calculated in QMC can be compared to this effective field-theoretic description.  
In addition, an associated critical point separating an XY-ferromagnet and a QSL can be demonstrated to be of an exotic ``fractionalized'' type, based on comparisons of 
universal quantities calculated from QMC and a related quantum field theory.

\subsection{BFG Hamiltonians with Z$_2$ spin liquid phases}
\label{sec:bfgz2}

One successful recipe for designing Hamiltonians with QSL phases amenable to QMC study was pioneered by Isakov and co-workers for certain Bose-Hubbard models on the kagome 
lattice \cite{Isakov1, Isakov2, TopoEE}.  Models in this class, first proposed analytically by Balents, Fisher and Girvin \cite{BFG} (BFG), are constructed partly based on their 
relation to quantum dimer models (QDMs), where a strong local constraint (the number of dimers emanating from a site being fixed) is coupled with a quantum tunneling that allows 
the state to fluctuate, while keeping the constraint satisfied. QDM Hamiltonians can be constructed where the groundstate superposition of dimer configurations breaks no 
symmetry---being a type of QSL (or resonating valence bond) phase, that has an emergent gauge symmetry and related topological order \cite{Misguich1}.

\begin{figure}
\includegraphics[width=8cm, clip]{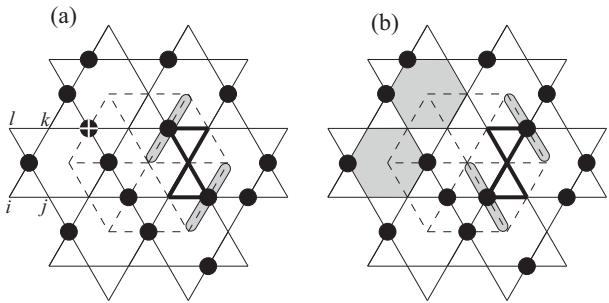}
\caption{ (a) Spins (black dots are $S^z = 1/2$, empty sites $S^z = -1/2$) in a groundstate configuration satisfying $H_0$, Eq.~(\ref{H0}), on the kagome lattice.  
The degenerate manifold of such configurations maps onto a triangular-lattice (dashed) classical three-dimer configuration, where each spin-up corresponds to a dimer 
on the dual lattice \cite{BFG}. (b) A plaquette-flip $H_{\rm ring}$ operation preserves the cluster-charging constraint, and can be seen to correspond 
to a rhombus-flip of dimers on the dual lattice.  A defect is created by flipping one spin, marked ``+'' in (a).  This defect can be thought of as associating with 
two hexagonal plaquettes, indicated in grey in (b), each with two spin-up per hexagon, corresponding to deconfined fractionalized spinons.} \label{kag_fig}
\end{figure}

The recipe to construct QSL phases in XXZ spin (or boson) models is to map the dimer constraint to a plaquette spin configuration on the dual lattice; the dimer may represent an $S^z$ spin configuration on a bond which is frustrated (or unsatisfied).  Then, the role of the quantum dimer tunneling moves is done by a spin-exchange processes (which may {\it not} be a simple two-body term).  Such models have been called
{\it cluster-charging} models \cite{Isakov2}, because spin configurations summed over a local cluster (e.g.~a lattice plaquette) are penalized when differing from some chosen value.  For example, on the kagome lattice (Fig.~\ref{kag_fig}),
\begin{eqnarray}
H_0 &=& V \sum_{\hexagon} (S^z_{\hexagon})^2,  \label{H0}
\end{eqnarray}
where 
\begin{equation}
S^z_{\hexagon} = \sum_{i \in \hexagon}S^z_i
\end{equation}
is a cluster-charging potential term which favors three spin up and three spins down per hexagonal plaquette.  This Hamiltonian alone will promote a highly degenerate manifold of groundstate configurations, where each kagome-lattice hexagon satisfies the plaquette constraint but is otherwise disordered.  
Like all operators which are entirely diagonal in the chosen QMC basis, $H_0$ does not affect the simulation with a sign problem, no matter how the interaction is frustrated, making it a suitable ingredient for a designer Hamiltonian.

In order to promote a QSL phase from this classical degenerate manifold, one desires to add quantum fluctuations to the Hamiltonian such that the cluster-charging constraint is not violated.  In this case, a {\it ring-exchange} term, 
\begin{eqnarray}
H_{\rm ring} = -K \sum_{\bowtie} ( S_i^{+} S_j^{-} S_k^{+} S_l^{-} + {\rm h.c.} ),
\end{eqnarray} 
which operates around a ``bow-tie'' plaquette on the kagome lattice (Fig.~\ref{kag_fig}),
promotes quantum fluctuations while preserving the cluster-charging constraint $H_0$.  Thus, one expects the kagome lattice model with $H = H_0 + H_{\rm ring}$ to retain a disordered groundstate with quantum fluctuations---a good candidate for a quantum spin liquid phase.  In addition, with $K>0$, such terms do not have the sign problem for QMC \cite{JKqmc}.

BFG \cite{BFG} were the first to show that variations of this kagome-lattice model support the simplest type of $Z_2$ spin liquid---in particular, at an exactly soluble Roksar-Kivelson point \cite{Rokhsar88}.  Subsequently, using QMC simulations, Isakov and co-workers showed that a variant of the cluster-charging Hamiltonian with different short-range spin exchanges, 
\begin{eqnarray}
H_{1,3} &=& -t_{1,3} \sum_{( ij )_{1,3}} [S^{+}_i S^-_j + S^-_i S^{+}_j],  
\end{eqnarray}
support a $Z_2$ spin liquid phase.  This exchange term can either connect the first, second and third neighbors $( ij )_3$ on the kagome lattice \cite{Isakov1,Isakov2}; or it can be nearest-neighbor only, $(ij)_1$ \cite{TopoEE}.  Again, for $t_{1,3}>0$ no sign problem exists. 
In addition to  $H = H_0 + H_3$ and $H = H_0 + H_1$, the $Z_2$ spin liquid has been demonstrated on $H = H_1 + H_{\rm ring}$, where the cluster-charging term is not explicitly required and the Hamiltonian consists of two competing kinetic-energy terms \cite{Long}, sometimes called the J-K model \cite{JKqmc}.

In the next section, we explore one procedure by which QMC simulations can positively identify such spin liquid phases.  In section \ref{XYstar}, we examine the XY-ferromagnet (or ``superfluid'' in the boson language) to QSL phase transition, which is common to these models, as a different example of a deconfined or fractionalized quantum critical point.

\subsection{Identifying the $Z_2$ spin liquid with entanglement} 
\label{topoEEsec}

The smoking gun signature for a gapped spin liquid phase could be argued to be the identification of an emergent gauge theory,
and the associated topological order \cite{Wenbook}.  One manifestation is the existence of a ground state degeneracy, which is topological in nature, in the simplest ($Z_2$) case being four-fold on a torus.  Previous QMC studies of QSL states have demonstrated the difficulty in clearly identifying this degeneracy, due to the tendency for tunneling between the equal-energy states on finite-size lattices \cite{Isakov1}.  Similarly, recent DMRG work studying candidate gapped spin liquid states have been unable to identify the expected topological degeneracy \cite{Yan,J1J2}.

\subsubsection{The boundary law}

Luckily, there is another tool suited for the positive identification of this topological order, through measurements of the Renyi entanglement entropy, introduced in section \ref{ss:renyi}.  The quantities $S_{\alpha}$ are well studied in quantum information science; they quantify correlations for a system in a basis-independent way, which makes them suitable in particular for characterizing phases that do not have an explicit broken symmetry (i.e., QSL phases).  In fact, some authors have championed the entanglement entropy as a paradigmatic analog to symmetry breaking for non-Landau phases; here, the concept of {\it Long Range Entanglement} (LRE) replaces the concept of {\it Long Range Order} (LRO) \cite{Wenbook}.  For example, in a gapped phase with LRO (such as the VBS phases mentioned above), the entanglement entropy is expected to obey the ``boundary law'' (also called the ``area law'')
\begin{equation} 
S_{\alpha} = a\ell + \cdots \label{arealaw}
\end{equation}
where $a$ is a non-universal constant, $\ell$ is the length of the boundary between regions $A$ and $B$, and terms denoted by $\cdots$ scale away at least as fast as $\mathcal{O} (1/\ell)$.  This can heuristically be considered as ``short-range'' entanglement.  In contrast, in a gapped spin liquid phase (with no broken symmetries)
scaling of entanglement entropy is predicted to be of the form
\begin{equation}
S_{\alpha} = a \ell - \gamma + \cdots \label{areaL}.
\end{equation}
 Here, the subleading term to the boundary law is a universal correction called the {\it topological entanglement entropy} \cite{Alioscia1,Alioscia2,LW,KP}.  In a gapped quantum spin liquid phase, it is related to the emergent gauge symmetry, independent of the Renyi index $\alpha$ \cite{Flammia}.  In particular, in the $Z_2$ spin liquid state, $\gamma =  \ln(2)$ at $T=0$ \cite{LW}.   The addition of this universal subleading constant to the boundary law motivates the notion of LRE for a gapped quantum spin liquid phase.

\subsubsection{Calculating the topological entanglement entropy}

Using the QMC estimators described in Section~\ref{ss:renyi}, it is possible to isolate the topological entanglement entropy from the leading-order boundary law, as well as subleading corrections due to effects such as corners, by taking measurements using several different geometries $A$ and performing additions or subtractions which isolate only $\gamma$.  For example, the {\it Levin-Wen} \cite{LW} construction isolates $2\gamma$ from Eq.~(\ref{areaL}), requiring measurement of four unique region $A$ geometries.  For a BFG Hamiltonian discussed in the 
previous section, it is possible to extract the topological entanglement entropy in a QMC simulation using this construction.  As noted by Castelnovo and Chamon \cite{castelnovo} through direct calculation on the toric code (which realizes a $Z_2$ fractionalized spin liquid groundstate), the value of $2\gamma$ is approached through two temperature crossovers on a finite-size system, related to the energy of the quasiparticle excitations in the groundstate---a $Z_2$ charge corresponding to the spinon, and a $Z_2$ flux corresponding to the vison.  Each excitation contributes a value of $\ln(2)$ to the topological entanglement entropy.  As illustrated in Fig.~\ref{QSLfig}, the realization of one of these plateaus at finite-temperature gives positive identification that the groundstate of the model is indeed in a topological $Z_2$ QSL phase.

\subsection{Deconfinement at the XY$^*$ transition} 
\label{XYstar}

The fractional spinons of the $Z_2$ spin liquid phase (illustrated in Fig.~\ref{kag_fig}) can be thought of as a ``square-root'' operator of the physical spins.  Employing the mapping from $S=1/2$ spins to hard-core bosons, we can regard the physical spin operator $S^+$ or (equivalently) boson operator $b^{\dagger}=\phi^\dagger\phi^\dagger$ as being composed of two creation
operators $\phi^\dagger$ for the fractionalized spinon particles.  

\subsubsection{The anomalous dimension}

Surprisingly, it is predicted that the quantum phase transition between the XY-ferromagnet 
and the $Z_2$ spin liquid discussed in the last section can actually be mediated not by the physical bosons (which would undergo a quantum phase transition in the 3D XY universality class), but by the fractional spinon fields \cite{XYstar1,XYstar2,earlyXYstar}.  This leads to the same exponents $z$ and $\nu$ as in the ordinary $XY$ critical point.  However, the exponent $\eta$ controlling the equal time correlation function $\langle b^\dagger(0) b(x) \rangle$ 
is modified. In a way similar to the DQC point discussed in Sec.~\ref{ss:j1j2N}, $\eta=1$ in the $N\to 1$ limit and is expected to be anomalously large also for $N=1$.

Previous estimates for $\eta$ were obtained from field-theoretic treatments. From an $1/N$ expansion, Ref.~\cite{XYstar2} obtains
\begin{equation}
\eta = 1 + \frac{32}{3 \pi^2 N},
\end{equation}
which gives $\eta = 2.08$ for $N=1$.
A more accurate value was obtained from a combination of field theory and Monte Carlo simulations of the correlation of a composite operator in the 3D XY 
model \cite{compositefieldtheory,compositeMC}, leading to $\eta\approx  1.47(3)$.

\begin{figure}
\includegraphics[width=8cm, clip]{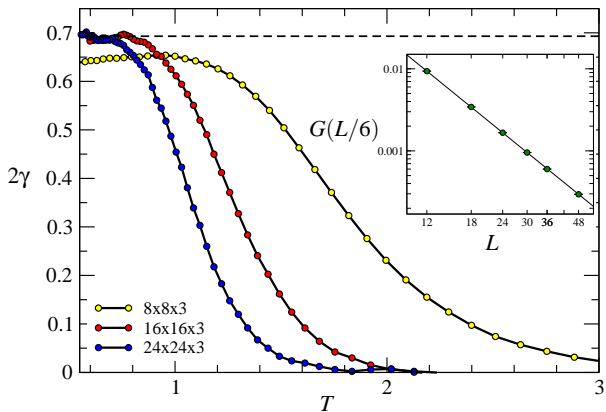}
\caption{Main graph: Temperature dependence of the topological entanglement entropy in the $Z_2$ QSL phase ($V/t_1 = 8$) of the Hamiltonian $H_0 + H_1$ 
for three different $L \times L \times 3$ kagome lattices \cite{TopoEE}.  The quantization of the plateau at $\ln(2)$ (dashed line) identifies the emergent 
gauge symmetry; one expect another temperature crossover to $2 \gamma = 2 \ln(2)$ at some lower temperature. Inset: finite-size scaling of $G(L/6)$ at the 
XY$^*$ quantum critical point ($V/t_1 = 7.0665$), which scales as  $1/L^{2.493}$ (straight-line fit) giving $\eta = 1.493$. Reproduced from Ref.~\cite{XYstarQMC}.}
\label{QSLfig}
\end{figure}

The measurement of $\eta$ in the kagome BFG model is challenging, requiring measurement the equal time Green's function in real space: 
$G(x)\equiv G(\tau=0,r) = \langle b^\dagger(0) b(r) \rangle$ \cite{WormA,gfsse}, done by keeping track of the defects created in the SSE directed-loop algorithm \cite{Syljuasen02} 
as it traverses the $(d+1)$-dimensional space-time QMC simulation cell.  At the critical point, $G(r)$ should decay as $1/r^{1+\eta}$, and finite size effects can be minimized, e.g., by measuring $G(L/6)$, which decays as $1/L^{1+\eta}$, as a function of $L$.  Fig.~\ref{QSLfig} show an algebraic decay with $\eta=1.493(10)$; a value consistent with $\eta$ for the composite operator in the 3D XY field theory mentioned above.  The excellent agreement between the QMC simulation results and the composite-boson field theory is a remarkable confirmation of fractionalized universality in this model; synergetic work between QMC and quantum field theory is uniquely capable of positive identification of the exotic quantum phase transitions in this model.

\subsubsection{Detecting fractionalization}

As discussed in Ref.~\cite{XYstarQMC}, there are consequences of this novel fractionalization in other universal quantities.  Remarkably, even though the fractional spinons become gapless at the quantum critical point, there are still observable remnants of the topological nature of the $Z_2$ spin liquid.  This can be seen by looking at the winding number distribution at the 
XY$^*$ point, and comparing it to the winding number distribution at a regular 3D XY transition.  Namely, since the fractional spinons are bosonic, they can be modeled by a ``simulation within a simulation'', achieved by projecting physical bosons at an XY transition to wind strictly in pairs around the space- and time-periodic simulation cell.  Since this distribution is universal, it is expected to be the same (up to a non-universal velocity) as the winding number distribution of the physical (composite) bosons in the XY$^*$ transition.  Observation of this fact \cite{XYstarQMC} is a striking confirmation that topological properties can still be present even when fractional excitations are gapless.  These universal winding number distributions are possible to compute in principle in field theory; hence they are another example of the synergetic connection between and quantum field theory and QMC simulations on designer Hamiltonians.

\section{Discussion}
\label{sec:discussion}

We have presented several examples of the use of judiciously constructed designer Hamiltonians for testing field theories
and exploring various quantum many-body phenomena related to quantum-criticality and exotic ground states. Here we discuss some further 
generic aspects of this approach, followed by a brief discussion of other interesting related studies of systems not covered in this 
Review, and an outlook on future directions.

\subsection{Designer Hamiltonian, stoquasticity, and the sign problem}

The concept of designer Hamiltonians that we have introduced here is different from that of ``stoquastic'' Hamiltonians \cite{Terhal08}, which are
those for which all off-diagonal matrix elements are non-positive (i.e., sign-problem free) in the standard basis.  This is normally the  $z$-component 
basis for spin systems, although a broader definition with respect to an arbitrary basis has also been presented \cite{Terhal09}, which would seem to 
indicate that any Hamiltonian could in principle be cast into the stoquastic class. Our criterion is more one of scientific utility than a strict mathematical 
definition---models that are free from sign problems either automatically (in a practically useful basis), or through some 
other way of circumventing it (e.g., the Meron algorithm \cite{Chandrasekharan99} or determinant-based QMC of particle-hole symmetric Hubbard 
models \cite{White89,Assaad05,Assaad07}). To be considered a designer Hamiltonian, such a model should also represent a prototypical case of 
some interesting quantum-man body phenomenon which is difficult to access in an unbiased manner in other ways, i.e., the model is designed to shed
light on this problem (and must be ``de-signed'' to be practically useful for studies on large scale).

\subsubsection{Limitations of designer Hamiltonians}

An important fundamental problem is whether it is always possible to construct such designer Hamiltonians, or whether certain types of
states are beyond reach in practice because their physical properties are fundamentally tied to the difficulties in circumventing the sign problem.
For instance, the``Bose metal'' spin-liquid phase with a Fermi sea of emergent fermions appears to have a complex sign structure that  would be difficult
to reproduce in a sign-free Hamiltonian in a simple basis~\cite{motrunich2007:dbl}. On the other hand, as we have reviewed in Sec.~\ref{sec:bfgz2}, $Z_2$ spin-liquid
ground states have been found in models with U(1) symmetry. An interesting open challenge is to find  the $Z_2$ spin liquid (which itself is a broad class
if states \cite{Wen03}) in sign-free SU($2$) invariant models. It is certainly possible that these states cannot be realized with sign-problem free Hamiltonians
in a simple basis, but one can still not exclude that designer Hamiltonians can be constructed for which some useful basis can be found, or for which the
sign problem can be solved in some other way. Indeed, the recent discovery of a spin liquid in the Hubbard model on the honeycomb lattice \cite{Meng10} seems
to provide an example, although it remains to be seen whether this state really is a $Z_2$ topological state.

\subsubsection{Computational complexity}

Regarding the sign problem itself more broadly, Ref.~\cite{Troyer05} presented a proof that the sign problem cannot be solved in general. 
The problem here was defined in terms of constructing a QMC algorithm for which the computational effort scales as a power-law in the system
size. A particular problem is considered for which the removal of the negative signs still leaves a problem for which the Monte Carlo sampling is believed 
to scale exponentially in the system size and, thus, the sign problem was not solved, only recast into a different form. However, the model system chosen in
this demonstration was that of a spin glass, and the conclusion reached is equivalent to the statement that the quantum spin glass is no easier 
to solve than the corresponding classical spin glass. This is hardly a surprising result. The proof does not address QMC solutions of systems that 
are not directly associated with the well known difficulties of simulating classical frustrated ``glassy'' systems with complicated energy 
landscapes. Whether or not the sign problem can be circumvented (at least for representative designer Hamiltonians) for translationally invariant 
systems such as fermionic Hubbard models without particle-hole symmetry and SU($2$) frustrated quantum spins models remains an important open question.

Some non-trivial solutions of sign problems are worth mentioning.
In some cases, local basis transformations can be used to render all off diagonal matrix elements non-positive \cite{Nakamura97}.
One can some times apply more sophisticated ways to circumvent 
the problem, e.g., with the Meron algorithm \cite{Chandrasekharan99} or the newly discovered fermion bag approach~\cite{chandrasekharan2010:fbag}.
It remains to be seen whether these exiting developments can be applied to a wider range of fermionic and frustrated designer Hamiltonians.

\subsection{Other systems and future prospects}

\subsubsection{Spin models}

The models and quantum states we have discussed here only represent a small sample of recent research into fascinating topics in quantum magnetism. Other 
noteworthy examples include systems exhibiting Bose-Einstein condensation of magnons \cite{Nohadani04}, spin supersolids \cite{Sengupta07}, qudrupolar order 
\cite{Harada02b,Harada07}, and quantum compass models \cite{Wenzel08b}. Here as well, QMC studies of designer Hamiltonians played a crucial role 
in uncovering novel physics, and unresolved issues will require the kind of synergistic interplay between low-energy theories and QMC simulations that 
we have emphasized in this Review.

Within the classes of spin models we have discussed here, although the essential physics now has been ascertained, there still remains
important work to be done to improve on the quantitative aspects, e.g., the critical exponents. In the case of the studies aiming to connect
SU($N$) QMC simulations with large-$N$ expansions, summarized in Sec.~\ref{ss:j1j2q} and Fig.~\ref{fig:exp}, it would be very useful to go to still larger lattices, 
to check the convergence of the computed exponents more precisely. This is important, in particular, in light of the fact there there are significant scaling 
corrections in quantities such as the spin stiffness for $N=2$. Such corrections could potentially also affect the fitting procedures underlying 
the exponents graphed in Fig.~\ref{fig:exp} for small $N$. When higher-order $1/N$ analytical results eventually become available, it will be important to know 
the values of the exponents without possible remaining effects of scaling corrections. What is needed here is high-accuracy (small error bars) 
results for the correlation functions on large lattices. It will also be very interesting to use the methods discussed in Sec.~\ref{ss:renyi} to quantify 
the degree of entanglement in the SU($2$) and SU($N$) models.

\subsubsection{Fermions}

Another very interesting aspect of the designer Hamiltonians we have discussed here is that they can also be doped with fermions. While this 
introduces a sign problem, it should be possible to study large lattices doped with a small number of holes, since the magnitude of the 
sign problem only depends on the number of fermions. This low-density limit of the SU($N$) models should already be sufficient for making connections 
with large-$N$ theories of ``holonic'' or ``strange'' metallic states, that have been predicted to arise out of the DQC system upon 
doping \cite{kaul2008:u1}.

An interesting conceptual question is whether one can study bosonic systems in which fermions are emergent particles.
A positive answer may be suggested by the demonstrated existence of the $Z_2$ fractional phase in a purely bosonic model, as we
discussed here in Sec.~\ref{topoEEsec}. 
The finite-temperature structure of the topological entanglement entropy identifies that both a $Z_2$ charge (a spinon) and a $Z_2$ flux (a vison) are realized as excitations in this model \cite{TopoEE}.
This raises the possibility that, with a suitable ``binding'' potential \cite{FermionBind} coupling these two, an extension of 
the model with an emergent fermionic excitation might be created \cite{Wenbook}.  
The ability for QMC to measure the entanglement entropy in a state with an emergent fermion
could also have wide implications for the purported application of tensor network algorithms as a ``solution'' to the sign problem, 
if such a state were found to violate the boundary law, Eq.~(\ref{arealaw}). 

\subsubsection{Disordered systems and impurities}

In this Review we have focused on translationally invariant systems. Systems with disorder, i.e., random couplings, dilution, etc., is another area where designer 
Hamiltonians can provide unique insights. To mention one example of a spin system for which the standard classical--quantum correspondence fails, in the 2D $S=1/2$ 
Heisenberg antiferromagnet diluted with non-magnetic impurities, the interplay of classical percolation and quantum fluctuations lead to low-energy 
excitations \cite{Wang10} not captured \cite{Vojta05} within the conventional $d$ $\to$ $(d+1)$-dimensional mapping. Other unanticipated effects of dilution 
on quantum-critical scaling have also been observed that still lack explanation and deserve further study \cite{Yu06,Roscilde06,Sandvik06,Roscilde07,Yao10}. 
For the designer Hamiltonians discussed in this Review, disorder effects have not yet been studied. One can anticipate interesting behaviors in VBS states, 
at the N\'eel--VBS transition, and in the Z$_2$ spin liquids. 

Beyond disordered spin models, various manifestations of the ``dirty boson'' problem remain intriguing and challenging \cite{Pollet09,Meier12,Iyer12}, and provide 
excellent examples of the indispensable role of QMC studies of designer Hamiltonians \cite{Gurarie09}. An example is the extended hardcore Bose-Hubbard model with 
random frustrated interactions on a 3D cubic lattice. QMC simulations have been instrumental in discovering that glassiness can coexist with superfluidity, 
and that quantum fluctuations and random frustration are both crucial ingredients needed to stabilize a ``superglass'' state \cite{superglass1,superglass2}.

In addition to extensive disorder, single-impurity problems are also interesting and important. Effects of a single missing or added spin have been
investigated with QMC in the 2D and 3D N\'eel states \cite{Hoglund04,Anfuso06,Liu09} and at conventional quantum-critical points in dimerized models \cite{Hoglund07}. 
QMC studies have also been done on vacancies in the VBS state of the 2D $J$-$Q$ model \cite{Kaul08}, where one can observe a vortex in the order parameter, 
and at the SU($2$) and SU($3$) critical points \cite{banerjee2010:log,banerjee2010:su3}, where impurities provide an interesting way to probe the anomalously 
large corrections to scaling in these systems. Impurities in 1D chains (with open edges) have also been studied \cite{Sanyal11}.

In general, impurities and disorder in interacting quantum systems present rich opportunities for further exploring exotic states and phase transitions.
Synergistic combinations of low-energy field theories, RG schemes for impurities \cite{Vojta12} and disorder \cite{Hoyos08,Vojta10,Altman10,Iyer12}, 
and QMC studies of designer Hamiltonians will be needed to make progress here.

\subsubsection{Non-equilibrium dynamic scaling}

A recent development on the methods front is the adaptation of QMC algorithms such as those discussed in Sec.~\ref{sec:methods} to non-equilibrium 
setups \cite{Degrandi11}. By studying time evolution out of a non-equilibrium state in imaginary time, standard sign-problem free systems
can be analyzed in novel ways. For instance, it was found in Ref.~\cite{Degrandi11} that the dynamic scaling properties when quenching to
a quantum-critical point is the same in real and imaginary time, and that such protocols are useful for extracting the dynamic critical
exponent. Moreover, the fidelity and full geometric tensor, which characterize the geometrical properties of the state space, can be
accessed this way. One can anticipate many fruitful applications of this unconventional QMC approach, which gives access to dynamics beyond the commonly
used numerical analytic continuation of QMC-computed equilibrium imaginary-timecorrelation functions \cite{Syljuasen08}.

\section*{Acknowledgments}

This work was supported in part by the NSF under Grants DMR-1056536~(RKK) and DMR-1104708~(AWS), and by the Natural Sciences and Engineering Research Council of Canada (RGM).
We also gratefully acknowledge support from the NSF I2CAM International Materials Institute Award, Grant DMR-0844115, as well as the International Center for Theoretical 
Physics (ICTP), Trieste, Italy, for support for the workshop {\it Synergies between Field Theory and Exact Computational Methods in Strongly Correlated Quantum Matter}, 
at ICTP in July 2011, where the foundations for this article were laid.

\end{document}